# Charge Transport in Doped Conjugated Polymers for Organic Thermoelectrics

Dorothea Scheunemann[1,2]*, Emmy Järsvall[1], Jian Liu[1], Davide Beretta[3], Simone Fabiano[4], Mario Caironi[5], Martijn Kemerink[2,6], Christian Müller[1]*

[1] Department of Chemistry and Chemical Engineering, Chalmers University of Technology, 41296 Göteborg, Sweden

[2] Centre for Advanced Materials, Heidelberg University, Im Neuenheimer Feld 225, 69120 Heidelberg, Germany

[3] Empa, Swiss Federal Laboratories for Materials Science and Technology, Ueberlandstrasse 129, 8600 Dübendorf, Switzerland

[4] Laboratory of Organic Electronics, Department of Science and Technology, Linköping University, Norrköping, Sweden

[5] Center for Nano Science and Technology@PoliMi, Istituto Italiano di Tecnologia, Via Pascoli 70/3, Milano 20133, Italy

[6] Department of Physics, Chemistry and Biology, Linköping University, Linköping, Sweden.

* dorothea.scheunemann@cam.uni-heidelberg.de; christian.muller@chalmers.se

## ABSTRACT

Research on conjugated polymers for thermoelectric applications has made tremendous progress in recent years, which is accompanied by surging interest in molecular doping as a means to achieve the high electrical conductivities that are required. A detailed understanding of the complex relationship between the doping process, the structural as well as energetic properties of the polymer films, and the resulting thermoelectric behavior is slowly emerging. This review summarizes recent developments and strategies that permit enhancing the electrical conductivity of p- and n-type conjugated polymers via molecular doping. The impact of the chemical design of both the polymer and the dopant, the processing conditions, and the resulting nanostructure on the doping efficiency and stability of the doped state are discussed. Attention is paid to the interdependence of the electrical and thermal transport characteristics of semiconductor host-dopant systems as well as the Seebeck coefficient. Strategies that permit to improve the thermoelectric performance, such as uniaxial alignment of the polymer backbone, both in bulk and thin film geometries, manipulation of the dielectric constant of the polymer, and variation of the dopant size are explored. A combination of theory and experiment is predicted to yield new chemical design principles and processing schemes that will ultimately give rise to the next generation of organic thermoelectric materials.

1. Introduction

Thermoelectric materials enable direct conversion between heat and electricity by converting a temperature difference into an electrical potential via the so-called Seebeck effect.[1] The energy conversion efficiency of a thermoelectric material can be described by a dimensionless figure of merit:

$$zT = \frac{\alpha^2 \sigma T}{\kappa} \tag{1.1}$$

where α is the Seebeck coefficient, $\sigma$ is the electrical conductivity, T is the absolute temperature, and $\kappa$ is the thermal conductivity. Instead of $zT$, the power factor $\alpha^2\sigma$ is often used to evaluate the performance of different thermoelectric materials.

Both the electrical conductivity as well as the Seebeck coefficient are related to the density of states (DOS), denoted as $g(E)$, of the system and the location of the Fermi level $E_F$. Following the derivation of Fritzsche,[2] the Seebeck coefficient of a semiconductor can be defined as the average entropy per charge carrier weighted by its contribution to the total electrical conductivity:

$$\alpha = -\frac{k_\text{B}}{q} \int dE \left(\frac{E - E_F}{k_\text{B}T}\right)\left(\frac{-\partial f}{\partial E}\right)\frac{\sigma'(E)}{\sigma} \tag{1.2}$$

where $k_B$ is the Boltzmann constant, $f$ is the Fermi distribution function, $q$ is the elementary charge and $\sigma'(E)$ is the energy dependent conductivity distribution function which is related to the electrical conductivity via:

$$\sigma = \int dE \left(\frac{-\partial f}{\partial E}\right)\sigma'(E) \tag{1.3}$$

with $\sigma'(E) = q \cdot g(E)\, f(E)\, \mu(E)$ and $\mu(E)$ being the energy dependent mobility distribution. As both $\alpha$ and $\sigma$ depend on the total carrier concentration $N = \int g(E)f(E)dE$, a common representation is to plot these two parameters against each other to circumvent the often-encountered difficulty of determining the total carrier concentration. When doing so a power law scaling with:

$$\alpha \propto \sigma^m \tag{1.4}$$

as shown in Fig. 1(a) is often observed. By collecting data for a wide range of materials, Glaudell et. al suggested that $m = -1/4$.[3] Such a scaling law would imply that the power factor scales according to $\alpha^2\sigma \propto \sigma^{1/2}$ (see Fig. 1b). However, significant deviations from this behavior can be found and we discuss these relationships in detail in section 5.

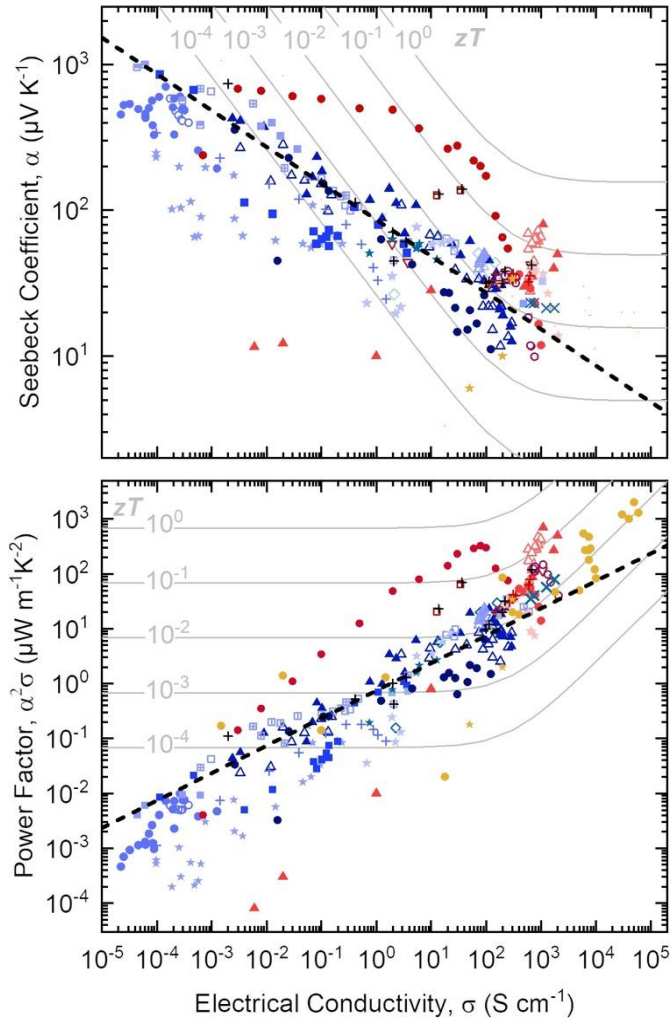

**FIG. 1. Interplay of thermoelectric parameters:** Seebeck coefficient $\alpha$ (top) and power factor $\alpha^2\sigma$ (bottom) as a function of electrical conductivity σ showing a universal trend. Data for p-type conducting polymers (blue: polythiophenes, red: poly(3,4-ethylenedioxythiophene) (PEDOT) based materials, yellow: polyacetylene) were taken from references listed in ref. 4; the dashed lines were constructed using equation 1.4 with $m = -1/4$; grey isolines corresponding to different figures of merit $zT$ were drawn using equation 6.4 with $T$ = 300 K and $\kappa_{ph}$ = 0.2 W m$^{-1}$ K$^{-1}$. Reproduced with permission from Mater. Sci. Eng. R Rep. 138 (2019). Copyright 2019 Elsevier.

In consequence, many authors assume that for an organic thermoelectric material the power factor increases with its electrical conductivity,[3, 4] and hence research currently focuses on maximizing the latter. As discussed above, the electrical conductivity depends on both the mobility and the carrier concentration. However, in case of conjugated polymers the intrinsic carrier concentration is generally low. An effective way to tune the carrier concentration and conductivity of conjugated polymers over several orders of magnitude, i.e. from $10^{-5}$ S cm$^{-1}$ (e.g. oxygen-doped poly(3-hexylthiophene) (P3HT)[5]) to $10^{4}$-$10^{5}$ S cm$^{-1}$ (polyacetylene,[6, 7] rubbed PBTTT[8]), is molecular doping.

How exactly molecular doping influences the electrical conductivity as well as other thermoelectric parameters, and thus ultimately the figure of merit, is however challenging to predict. On the one hand, several factors such as the energy landscape of the host-dopant system, its dielectric properties and the resulting nanostructure influence transport processes and the Seebeck coefficient. On the other hand, the parameters that determine $zT$ are interdependent and optimizing one of them tends to compromise the others. As a result, there appears to be an optimal range of about $\sigma \approx 10^2$ to $10^3$ S cm$^{-1}$ with respect to $zT$ [cf. electronic contribution to the thermal conductivity in section 6], as can be seen from the iso-$zT$ lines in Fig. 1b. Moreover, each of the thermoelectric parameters is temperature dependent and hence $zT$ will also vary with temperature.

This review focuses on the impact of doping on various parameters affecting the electrical conductivity, but also the Seebeck coefficient and the thermal conductivity of conjugated polymers. The chemical structures of all polymers and dopants discussed in this review are depicted in Fig. 2.

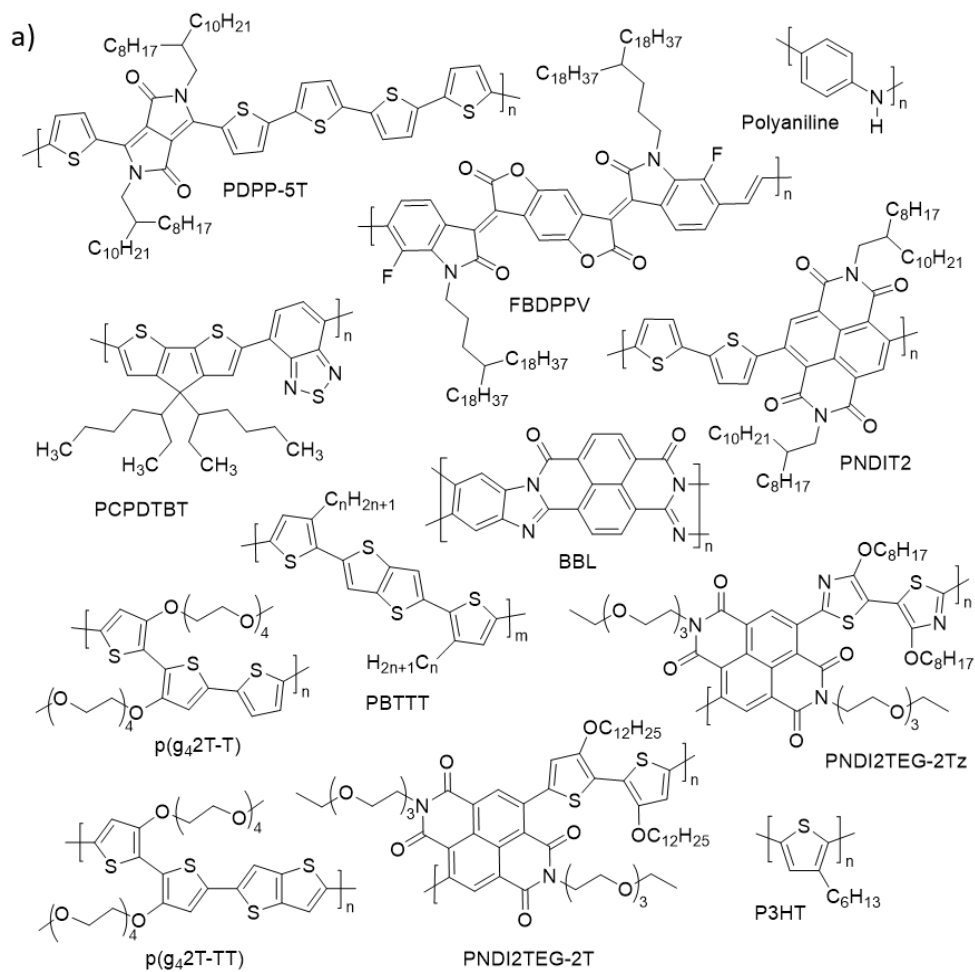
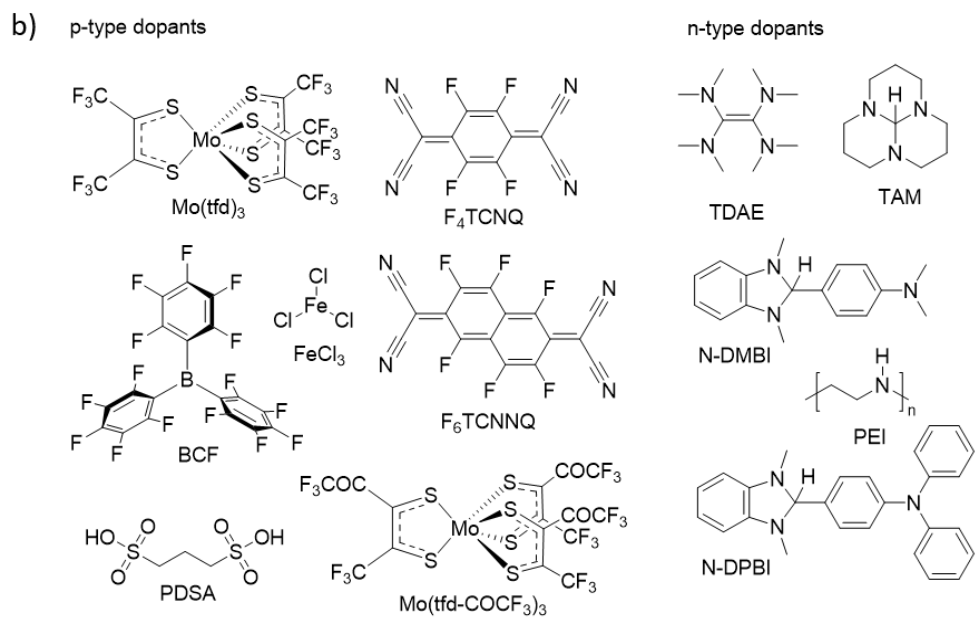

**FIG. 2.** Chemical structures of (a) conjugated polymers and (b) dopants discussed within this review.

We would like to point out that there are other types of materials, notably poly(3,4-ethylenedioxythiophene):poly(styrenesulfonate) (PEDOT:PSS) and poly(3,4-ethylenedioxythiophene):tosylate (PEDOT:Tos), where the conducting form of the polymer is created through oxidative polymerization,[9] resulting in p-type conductors that can display a power factor $\alpha^2\sigma$ >> 100 μW m$^{-1}$ K$^{-2}$ and a figure of merit $zT > 0.2$.[10-12] We refer the reader to recent reviews that discuss PEDOT-based materials in detail.[13-15]

Here, we focus on molecularly doped materials whose thermoelectric properties can be widely tuned through chemical design of both the polymer as well as the dopant molecules.

## 2. Molecular Doping

Molecular doping of conjugated polymers involves the addition of a chemical species that can undergo charge transfer with the host material. Many dopants can directly exchange a charge with the polymer. Redox active molecules are capable of either accepting or donating an electron from/to the polymer, which leads to p- and n-doping, respectively. Instead, the charge can also be transferred in the form of a proton (H$^+$) or hydride (H$^-$): some polymers such as polyaniline[16] and polythiophenes[17] can accept a proton from an acid and hence become p-doped while some semiconductors such as fullerenes[18] and amide- or ester-rigidified bis(styryl)benzene derivatives[19] become n-doped when they receive a hydride from DMBI-based dopants (see Fig. 2 for chemical structures).[18, 20] A similar mechanism has been proposed for, e.g., the naphthalenediimide (NDI)-based copolymer PNDIT2 [see Fig. 2],[18, 20] although never confirmed experimentally. It was instead argued that for weak hydride acceptors such as PNDIT2, a thermally-activated homolytic C–H bond cleavage (C–H → C$^\cdot$ + H$^\cdot$) has a more significant impact.[21] Recently, the use of transition metals of the group 8-11 elements has been shown to catalyze the heterolytic cleavage of the N-DMBI-H (C–H → C$^+$ + H$^-$) bond, enabling greatly increased doping efficiency.[22] Other p-doping mechanisms involve a second species such as atmospheric oxygen or water that either act as the actual oxidant or first react with the dopant to form an active compound that then undergoes charge transfer with the polymer. Examples include acid mediated oxidation through oxygen[23] and doping with the Lewis acid tris(pentafluorophenyl)borane (BCF), which first forms a complex with water that then donates a proton to the polymer, followed by an additional electron transfer with an adjacent polymer chain.[24, 25] The variety of doping mechanisms that can take place have been reviewed in detail.[26, 27] In the remainder of this review, we will focus on direct electron transfer, which is the mechanism that is encountered when using a wide range of molecular dopants.

### 2.1. Degree of Charge Transfer and Ionization Efficiency

A dopant and host can undergo integer or partial electron transfer and then form an ion pair or charge-transfer complex (CTC), respectively.[28, 29] In case of integer charge transfer the charge that is created on the polymer chain, referred to as a polaron, is balanced by the opposite charge on the dopant molecule, which now takes on the role of the counterion. The polaron and counterion form an ion pair that remains Coulombically bound. Not every dopant molecule that is introduced to the host experiences electron transfer, for example because of aggregation or variations in the energy levels of the host matrix (cf. discussion below). The ionization efficiency, which is typically less than 100%, is given by:

$$\eta_{\text{ion}} = \frac{N_{polaron}}{N_{dopant}} \tag{2.1}$$

where $N_{polaron}$ and $N_{dopant}$ are the number of generated polarons and dopant molecules per unit volume. While $N_{dopant}$ is known *a priori*, especially in case of co-processing [cf. section 3.6], at low to moderate doping levels $N_{polaron}$ can be estimated with a number of techniques such as electron paramagnetic resonance (EPR)[30] or optical spectroscopy.[25, 31-33] However, caution should be paid to changes in film texture or polaron delocalization with doping level, which can for example alter the infrared absorption spectrum of polarons.[34, 35]

Ion-pair formation is facilitated by a favorable offset between relevant energy levels of the two species. In case of p-doping the organic semiconductor donates an electron from its highest occupied molecular orbital (HOMO) to the lowest unoccupied molecular orbital (LUMO) of the (typically neutral) dopant, which readily occurs if the electron affinity of the latter ($EA_{dopant}$) is equal to or higher than the ionization energy of the former ($IE_{osc}$), i.e. $EA_{dopant} \geq IE_{osc}$ (Fig. 3(a)). For n-doping to occur, the opposite constellation is beneficial, i.e. the electron affinity of the organic semiconductor ($EA_{osc}$) should be equal to or larger than the ionization energy of the dopant ($IE_{dopant}$), i.e. $EA_{osc} \geq IE_{dopant}$ (Fig. 3(a)). The EA and IE of the dopant and host can be accurately determined with photoelectron spectroscopy or estimated using electrochemical methods such as cyclic voltammetry. However, values obtained for each individual species should only be used as an approximate guide to predict if electron transfer is energetically favorable since polarization and electrostatic interactions arise when placing the dopant in the host matrix, which can significantly alter the energy levels by up to 1 eV.[36] Besides, electrostatic interactions can be expected to stabilize the counterion-polaron pair[29] and hence may influence $\eta_{\text{ion}}$, which may be noticeable when comparing dopants with different sizes. Further, organic materials tend to show a high degree of structural and energetic disorder, which –as well as doping induced disorder– leads to a broadening of the DOS and hence allows electron transfer despite an unfavorable offset between the nominal energy levels measured with, e.g., electrochemical methods.[28, 37] For weak p-dopants with $EA_{dopant} < IE_{osc}$ broadening of the density of states can lead to an increase in $\eta_{ion}$ with the dopant concentration.[37, 38]

CTC formation, as schematically depicted in Fig. 3(b), involves hybridization of the frontier orbitals of the organic semiconductor and dopant and does not have any specific requirements on the energy levels of the two species. Hence, CTC formation can occur when ion-pair formation is not favored, e.g. if $EA_{dopant} < IE_{osc}$ in case of p-doping, but also when the energy offset allows the latter. For systems where CTC and ion-pair formation compete, aspects such as the nanostructure of the semiconductor[39, 40] and the degree of overlap of the frontier orbitals of the dopant and semiconductor[17, 41] determine which process dominates. While a CTC is neutral, it can in a subsequent step undergo electron transfer with an adjacent organic semiconductor site, and thus create a polaron, provided that its energy levels allow integer charge transfer.[42] Since doping via CTC formation is a two-step process the ionization efficiency is typically low, i.e. $\eta_{ion}$ << 100 %.

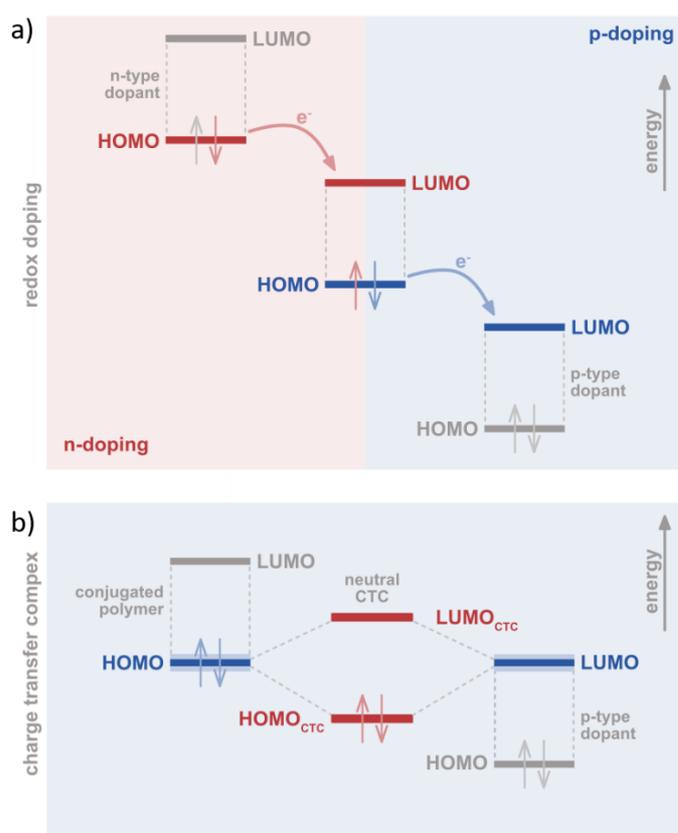

**FIG. 3**. **Influence of frontier orbitals on molecular doping processes:** (a) Basic principle of redox doping involving the transfer of an electron (e⁻) from the semiconductor HOMO to the dopant

LUMO in case of p-doping and (b) the formation of a charge-transfer complex through fractional charge transfer.

While many dopants are neutral species, electron transfer with the host can also be facilitated by radical cations[31, 43, 44] or anions[45] provided they have suitable energy levels. Accordingly, for some combinations of semiconductor and dopant, such as F$_4$TCNQ and polythiophenes with a sufficiently low IE, two electron transfer events can occur, where the initially neutral dopant molecule exchanges a first electron with the host followed by the exchange of a second electron between the generated counterion and a further host site.[45] This type of double doping can result in an ionization efficiency of up to 200%.

## 2.2. Dissociation

Polarons that form as a result of doping remain strongly bound to the counterion. These ion pairs must dissociate to generate free charges that can move in the host and contribute to charge transport. The dissociation efficiency is given by:

$$\eta_{diss} = \frac{N_{free}}{N_{polaron}} \tag{2.2}$$

where $N_{free}$ is the number of free charge carriers, which can be measured with admittance spectroscopy on metal-insulator-semiconductor (MIS) devices or AC-Hall measurements. We note that in case of materials with hopping-dominated transport such as doped conjugated polymers, the interpretation of AC Hall measurements is not entirely straightforward as the motion of hopping carriers can partially compensate the Hall voltage, which results in an overestimation of the carrier density.[46] The dissociation efficiency is typically much less than 100% because of the strong Coulomb binding energy between dopant and polaron, i.e. not every polaron can contribute to charge transport. The Coulomb binding energy $E_C$ is given by:

$$E_C(r) = \frac{q^2}{4\pi\varepsilon_0\varepsilon_r r} \tag{2.3}$$

where $q$ is the elementary charge, $\varepsilon_0$ the permittivity of vacuum, $\varepsilon_r$ the dielectric constant of the host and $r$ the distance between the polaron and the corresponding counterion. Assuming typical values for organic semiconductors such as $\varepsilon_r \approx 3$ and $r = 1$ nm, the binding energy is on the order of 500 meV, which is much larger than the thermal energy $k_B T$ = 25 meV at room temperature. Although the concept of free charges is somewhat difficult in a sytem with strong localization and a thermal capture radius $r_{\text{th}}$ that often is larger than the inter-dopant distance (for the given parameters, $r_{\text{th}} \approx 20$ nm, which corresponds to a relative dopant concentration of $c = 10^{-4}$), the dissociation efficiency of dopant-counterion pairs is typically argued to be low. For example, in case of P3HT doped with F$_4$TCNQ, which tend to undergo integer charge transfer, Pingel and Neher have shown that at low dopant levels only 5% of the generated ion pairs dissociate into free charge carriers and hence contribute to charge transport.[47] Molecular strategies that reduce Coulombic effects by increasing $\varepsilon_r$ or the polaron-counterion distance $r$ and thus improve $\eta_{diss}$ will be discussed in sections 3.3 and 3.4.

Depending on the doping level, the energetic landscape and electrostatic interactions can strongly reduce the activation energy for dissociation.[48, 49] At very low doping levels $\eta_{diss}$ is mostly governed by the Coulomb binding energy. At intermediate doping levels the energetic disorder that is introduced by neighbouring dopant molecules can strongly reduce the activation energy to as little as a few tens of meV,[50] which is comparable to $k_B T$ and hence thermal activation results in a higher $\eta_{diss}$ than what would be expected by only considering $E_C$. Moreover, Beljonne et al. investigated the role of interactions with the local environment on the generation of free charge carriers and found that molecular quadrupole moments of host and dopant molecules reduce the energy barrier for dissociation.[51] At higher doping levels, the Coulomb potentials of counterions and polarons significantly overlap and the electronic landscape flatens, which benefits charge transport [see section 3.2]. The concept of charge dissociation however becomes ambiguous at very high doping levels, where polarons are mostly trapped by the ubiquitous counterions, meaning there are nearly no free charge carriers. However, polarons can still contribute to charge transport after escaping from the Coulomb radius of one counterion until they are captured by another counterion. In that case, kinetic Monte Carlo (kMC) simulations are required to estimate the fraction of charges $f_{transport}$ that contributes to charge transport at any time to better evaluate the doping process.[52, 53]

### 2.3. Doping Efficiency

The overall doping process comprises two steps, ionization and dissociation [Fig. 4], and hence the doping efficiency is given by:

$$\eta_{doping} = \eta_{ion} \times \eta_{diss} = \frac{N_{free}}{N_{dopant}} \tag{2.4}$$

Hence, $\eta_{doping}$ is limited by factors that reduce $\eta_{ion}$, such as dopant aggregation and CTC formation [see section 2.1], as well as $\eta_{diss}$, such as strong Coulomb binding between polaron and counterion [see section 2.2].

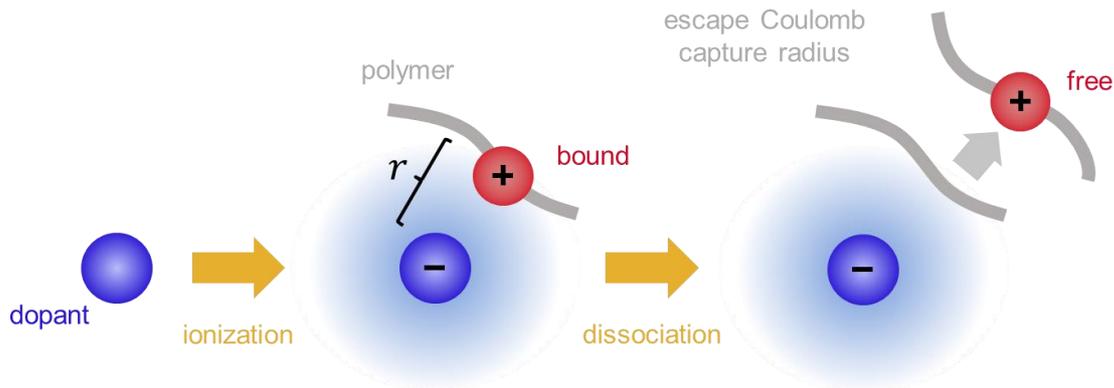

**FIG. 4.** Schematic illustration of the two steps of a doping process.

Since the thermoelectric performance of organic thermoelectric materials generally increases with the electrical conductivity [Fig. 1], it is important to maximize the *density* of charge carriers that contribute to transport $\rho_{transport}$, i.e. the doping level should be high. A large $N_{dopant}$ can however lead to aggregation of dopant molecules due to limited miscibility with the host, which reduces $\eta_{ion}$ and hence $\eta_{doping}$. The conducting material then consists of a fraction of dopant aggregates, which only occupy volume but do not significanlty contribute to doping, which reduces $\rho_{transport}$. Synthetic strategies that allow to adjust the polarity of the polymer and/or dopant and hence improve miscibility will be discussed in section 3.5. The energy barrier for dissociation at high doping levels can be influenced by electrical screening, e.g. by selecting a material with a high $\varepsilon_r$ [section 3.3], and by molecular design strategies that place the polaron and counterion further apart. It is important to keep in mind that a high $\rho_{transport}$ is needed, which means that the total volume of the conducting material constitutes an intrinsic limit. For example, a very large dopant molecule enhances $\eta_{diss}$ and hence $\eta_{doping}$ [cf. equation 2.4] but also occupies more space than a smaller dopant, leaving less volume that can be occupied by the charge-conducting conjugated material. Hence, there is likely an optimal dopant size for which $\rho_{transport}$ is maximized.

## 3. Parameters affecting charge transport and doping

### 3.1. Energy Offset

As discussed in section 2.1, a crucial aspect that influences doping through ion-pair formation is the energy offset $\Delta E$ between the IE and EA of the semiconductor host and dopant, ie. $\Delta E$ = EA$_{dopant}$ - IE$_{osc}$ in case of p-doping and $\Delta E$ = EA$_{osc}$ - IE$_{dopant}$ for n-doping. For systems based on doped polymers[54,55] as well as small molecules[37] it has been observed that the ionization efficiency increases with $\Delta E$. One challenge that is encountered when studying the effect of $\Delta E$ on the doping efficiency is the interplay of structural and energetic effects. For instance, the size of the dopant[56] and its miscibility with the polymer matrix[57] can significantly affect $\eta_{doping}$ [structural changes are discussed in sections 3.4 to 3.6]. Aubry et al. studied a molecular p-dopant system whose EA$_{dopant}$ can be tuned over a wide range from 4.9 to 5.7 eV while maintaining the dopant size and shape, which ensures that also the Coulomb interactions between polarons and counterions remain similar.[58] An increase in the EA of the p-dopant not only resulted in an increase in $\Delta E$ but also improved the redox-driven infiltration of the p-dopant during sequential doping, which, overall, led to a higher doping level. Likewise, Jacobs et al. observed that $\sigma$ increases with $\Delta E$ = EA$_{dopant}$ - IE$_{osc}$ for ion-exchange doping of PBTTT with a series of p-dopants with EA$_{dopant}$ ranging from 5.2 to 5.9 eV, likely because of an increase in $N_{polaron}$ and, possibly $\eta_{diss}$, which involved sequential doping by a molecular dopant that was dissolved in acetonitrile together with an excess amount of an ionic liquid, leading to ion-exchange of the dopant counterion and the anion of the ionic liquid subsequent to the doping reaction.[59] When discussing the influence of energetics on ionization of the semiconductor, it is important to keep in mind that electrostatic interactions can strongly alter the energy levels, as discussed in section 2.1. Further, the semiconductor matrix can display a complex nanostructure, for example disordered and ordered domains in a polymer film, with a locally changing IE$_{osc}$ and EA$_{osc}$. As a result, a certain dopant may only be able to undergo electron transfer with certain parts of a polymer film, e.g. the crystalline domains in polythiophenes, which have a lower IE$_{osc}$ than disordered regions.[60]

In a carefully optimized system, an $\eta_{ion}$ close to 100% can be achieved at low doping concentrations via co-processing or sequential processing, meaning that each dopant ion gives rise to one polaron.[54, 61] In case of high doping concentrations, however, many dopant molecules tend to aggregate, which leads to a reduction in $\eta_{ion}$, since now some of the dopant molecules do not undergo electron transfer with the semiconductor. The limit of $\eta_{ion}$ = 100 % can be surpassed through *double doping* if the $\Delta E$ is large, on the order of at least 0.5 eV. This is because dopants such as F$_4$TCNQ experience a reduction in electron affinity when they accept an electron, i.e. from EA$_{dopant}$ = 5.2 eV to EA$_{anion}$ = 4.7 eV. As a result, polythiophenes such as the thienothiophene-bithiophene copolymer p(g$_4$2T-TT) [see Fig. 2 for chemical structure], which features a low IE$_{osc}$ = 4.5 eV, can donate an electron to both neutral F$_4$TCNQ as well as its anion since EA$_{dopant}$ > EA$_{anion}$ > IE$_{osc}$.[45] Double doping permits an $\eta_{ion}$ of up to 200% and in case of low doping levels $\eta_{doping}$ of up to 200%. Tools that allow to reach a higher doping efficiency may allow to reduce the concentration of counterions in the semiconductor and therefore minimize the perturbation of the energetic landscape of the host [see section 3.5].

### 3.2. Density of States and Disorder

In intrinsic, undoped materials the density of states (DOS) is assumed to be (either exponential or) Gaussian in shape:

$$g_i(E) = \frac{N_0}{\sqrt{2\pi\sigma_{DOS}^2}} exp\left(\frac{-(E_i - E_0)^2}{2\sigma_{DOS}^2}\right) \tag{3.1}$$

where $E_i$ is the single particle energy of site $i$, $\sigma_{DOS}$ the energetic disorder of the Gaussian DOS, and $N_0$ is the total site density. Upon doping the shape of the DOS however changes as a consequence of the long range of the Coulomb potential of ionized dopants and the low dielectric constants of most organic semiconductors. An approximation for such an ion-perturbed DOS of a doped semiconductor at low to moderate doping levels, where Coulomb traps can be considered as independent, was developed by Arkhipov et al.,[62, 63] as:

$$g(E) = A \int_{-\infty}^{0} \frac{dE_C}{E_C^4} \exp\left(\frac{A}{3E_C^3}\right) g_i(E - E_C), \tag{3.2}$$

where $A = 4\pi q^6 N_{\text{dopant}}/(4\pi\varepsilon_0\varepsilon_r)^3$. Pingel and Neher used the resulting mobility model to fit the dependence of the electrical conductivity on the doping ratio of P3HT:F$_4$TCNQ films.[47]

Zuo et al. extended equation 3.2 to account for energy level differences between the dopant and the semiconductor, i.e. $\Delta E = \text{EA}_{\text{dopant}} - E_i$ in case of p-doping and $\Delta E = E_i - \text{IE}_{\text{dopant}}$:[5, 64]

$$\begin{aligned}
g(E) &= \left(1 - \frac{4\pi N_d}{3N_i}\right) \frac{g_1(E)}{\int_{-\infty}^{0} dE\, g_1(E)} + \frac{4\pi N_d}{3N_i} \frac{g_2(E)}{\int_{-\infty}^{0} dE\, g_2(E)} \\
g_1(E) &= A \int_{E_1}^{0} \frac{dE_C}{E_C^4} \exp\left(\frac{A}{3E_C^3}\right) g_i(E - E_C) \\
g_2(E) &= A \int_{-\infty}^{E_1} \frac{dE_C}{E_C^4} \exp\left(\frac{A}{3E_C^3}\right) g_i(E - \Delta E - E_C),
\end{aligned} \tag{3.3}$$

where $E_1 = E_C\left(N_i^{-1/3}\right)$ is the Coulomb energy one lattice constant away from the ionized dopant. The purpose of this extension is to construct a DOS, which contains weighted parts that reflect both the host material ($g_1$) and the dopant ($g_2$). It is worth noting that both Eq. (3.2) and (3.3) do neither consider contributions of non-nearest dopant ions to the Coulomb energy nor long-range Coulomb interactions between mobile charges. However, Zuo et al. demonstrated that the model presented in equation (3.3) shows, up to a relative dopant concentration of $10^{-2}$, good agreement with numerically accurate kMC simulations that include both effects. At high doping levels, however, these Coulomb interactions as well as the effect of double occupation on the DOS have to be taken into account in order to correctly describe the experimentally measured DOS.[65, 66] In addition, it was recently shown that, assuming an energy-dependent localization length, a semi-analytical model based on a Gaussian disorder model can qualitatively describe experimentally observed power-law trends of σ(*N*) even at medium to high doping concentrations. The energy dependence of the localization length was determined on the basis of numerically exact solutions of a tight binding model.[67]

Fig. 5(a) shows a Gaussian DOS broadened by Coulombic interactions as described by equation 3.3. Such a broadening of the DOS upon doping has been confirmed experimentally using photoelectron spectroscopy.[68-70]

As argued in section 2, polarons have to overcome the Coulomb energy barrier in order to contribute to charge transport. Several studies showed that electrostatic energetic disorder induced by neighboring counterions can facilitate this escape process.[5, 49, 50] At sufficiently high doping levels, the Coulomb potentials of individual counterions overlap [see Fig. 5(b)], which flattens the energy landscape and promotes the escape of carriers from long-range traps. Using kMC simulations Fediai et al. demonstrated that the intrinsic and doping-induced disorder correlate. Hence, the modification of the DOS is governed by disorder compensation effects.[71] For highly disordered materials the additional disorder introduced by doping can overcompensate the intrinsic disorder of the material. As a result, the total disorder remains constant or is even reduced.

An additional effect of doping induced disorder is that even nominally weak dopants can be almost fully ionized. In this context it was shown by kinetic Monte Carlo simulations that the ionization of weak dopants can be a disorder-activated process in organic materials, while it is temperature activated in inorganic semiconductors.[38] In turn this implies that the energy offset between the frontier orbitals of the dopant and host are less important than commonly assumed (section 3.1).

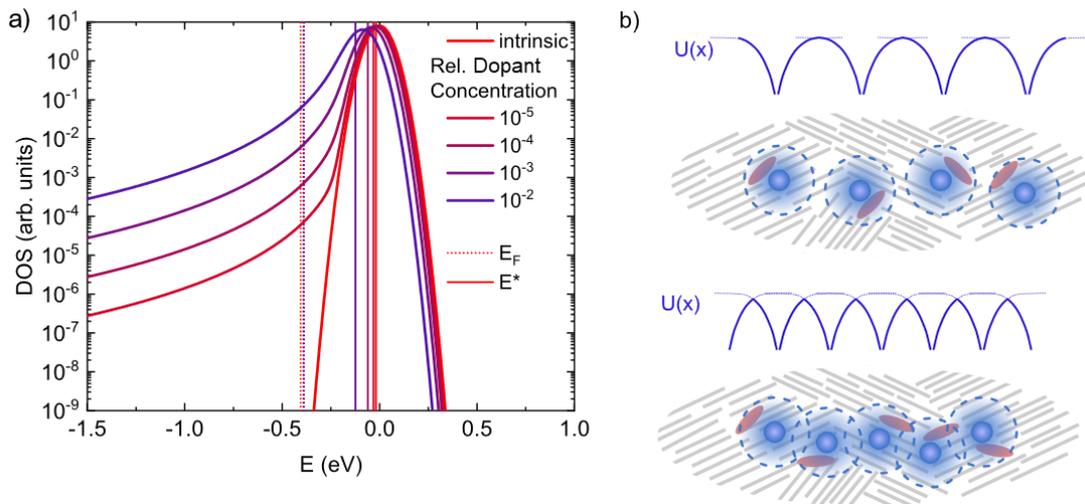

**FIG. 5. Impact of doping on the density of states (DOS):** Effect of low to moderate doping on a Gaussian DOS (energetic disorder σ$_{DOS}$ = 50 meV) of a disordered organic semiconductor (a) and effect of a high doping level on the Coulomb potential overlap (b). Thin dashed and solid lines in (a) indicate the Fermi Energy $E_f$ and the transport energy $E^*$, respectively. (b) (top) Negligible overlap of the Coulomb potentials $U(x)$ of counterions at a low doping level leads to large barriers ɸ for detrapping (bottom) while significant overlap at a sufficiently high doping level flattens the energy landscape.

### 3.3. Impact of the Dielectric Constant on Charge Transport

At high doping levels, which are needed for the design of thermoelectric materials, the charge on the polymer, i.e. the polaron, cannot move sufficiently far away from the dopant counterion to escape its Coulomb radius, without feeling the influence of another counterion [section 2.2]. Hence, the continuous interaction of polarons and surrounding dopant counterions (as well as other charges) must be considered. As a result, charge transport is dominated by the collective interactions between all charges within the conducting material. The extent to which a charge feels the presence of counterions in its vicinity depends, to a first approximation, on the dielectric constant of the *medium*. According to equation 2.3, an increase in dielectric constant will reduce the Coulomb binding energy that is felt by the polaron in the vicinity of a polaron, i.e. partial electrical screening results in a less strongly bound polaron. It can therefore be anticipated that the dielectric constant influences charge transport in doped conjugated polymers.

The dielectric constant of conjugated polymers can be adjusted through a variety of means. Strategies that are currently under investigation include partial fluorination of the polymer

15 / 48

backbone,[72-75] modification of non-polar alkyl side chains with polar sulfinyl, sulfonyl or cyano groups,[73, 76-78] and the replacement of non-polar side chains with more polar oligoether side chains.[79-83] Functional groups associated with strong local permanent dipoles may itself constitute a Coulomb trap. Cyano groups positioned at the end of alkyl side chains, for instance, increase the energetic disorder, which reduces the hole mobility deduced from space-charge limited current (SCLC) measurements by about two orders of magnitude.[73] The use of oligoether side chains appears to be a better choice for increasing the overall dielectric permittivity that is felt by a polaron traversing a doped polymeric material, since local variations in electric field are likely to be less pronounced. An additional advantage of oligoether side chains, as discussed in section 3.5, is the improved compatibility of polar polymers with dopant molecules.

A comparison of polythiophenes with alkyl or oligoether side chains, doped with the same molecular dopant, indicates that the latter tend to display a higher electrical conductivity. For example, doping of p(g$_4$2T-T)[84] with F$_4$TCNQ yields an electrical conductivity of up to 100 S cm$^{-1}$,[85] which is two to typically ten-fold higher than values reported for F$_4$TCNQ doped P3HT.[86-90] The polar tetraethylene glycol side chains of p(g$_4$2T-T) bestow the polymer with a higher $\varepsilon_r$ = 4.4 as compared to P3HT with $\varepsilon_r$ = 2.7.[91] The higher dielectric constant in case of p(g$_4$2T-T) can be expected to enhance screening of dopant anions such as F$_4$TCNQ (di)anions, which are thought to be located between the side chains of the polymer.[86, 91-95] As a result, it is feasible that polarons on the polythiophene backbone are less strongly influenced by the counterions surrounding the polymer, which would result in a higher charge-carrier mobility and hence electrical conductivity (cf. equation 1.3).

To support this argument, we vapor-doped thin films of P3HT and p(g$_4$2T-T) using a reported procedure[88] and determined the variable-temperature electrical conductivity $\sigma(T)$, which we fitted with a previously described analytical variable range hopping (VRH).[5, 64, 96] The good agreement of the experimental data and the VRH model, which takes the difference in $\varepsilon_r$ and oxidation level into account [Table 1 and Fig. 6], is consistent with the projected positive influence of a higher $\varepsilon_r$ on $\sigma(T)$. The size of $\varepsilon_r$ has an influence on the DOS as described in section 3.2. An increase of the dielectric permittivity decreases the Coulomb interactions and thus leads to a smaller width of the DOS. We anticipate that in-depth studies, which elucidate the impact of the local dielectric environment on charge transport, rather than treating the material as a dielectric continuum, will provide invaluable insights with regard to, e.g., the optimal choice of side chains. In this context, we would like to highlight the recent theoretical work by Comin et al., which shows that high doping levels lead to a strong increase in dielectric constant and large fluctuations at the microscopic scale.[97]

**Table 1.** Parameters obtained from fitting the variable-temperature conductivity $\sigma(T)$ of F$_4$TCNQ vapor doped P3HT and p(g$_4$2T-T) shown in Fig. 6, with the VRH model described in refs. [5, 64, 96], using fixed values of the dielectric constant $\varepsilon_r$ from ref. [91], and treating the attempt-to-hop frequency $v_0$, localization length of the wavefunction $\alpha^{-1}$, mean inter-site distance on a cubic lattice $a_{NN}$ and dopant-induced carrier density $N$ as free parameters. Here $N = c/a_{NN}^3$ with $c$ the relative dopant concentration assuming that each dopant produces one charge.

|  | **P3HT** | **p(g$_4$2T-T)** |
| --- | --- | --- |
| $\varepsilon_r$ | 2.7 | 4.4 |
| Gaussian disorder (meV) | 78 | 55 |
| $v_0$ (10$^{13}$ s$^{-1}$) | 2.7 | 9.5 |
| $\alpha^{-1}$ (nm) | 0.93 | 0.42 |
| $a_{NN}$ (nm) | 2.8 | 1.3 |
| $N$ (10$^{26}$ m$^{-3}$) | 0.2 | 0.6 |

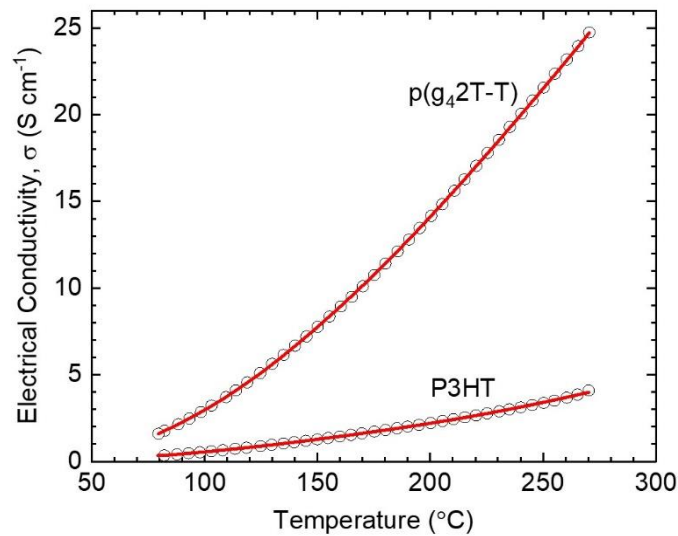

**FIG. 6. Impact of the dielectric constant on the electrical conductivity:** Variable-temperature electrical conductivity $\sigma$ (open symbols), and analytical fit using the VRH model described in refs. [5, 64, 96] with the fit parameters given in Table 1, and using experimentally determined values for $\varepsilon_r$ of P3HT (regioregularity $\approx$ 96 %, $M_n \approx$ 29 kg mol$^{-1}$; PDI $\approx$ 2.2; 70 nm thin film spin-coated from 1:1 chlorobenzene:o-dichlorobenzene at 60 °C) and p(g$_4$2T-T) ($M_n \approx$ 16 kg mol$^{-1}$; 85 nm thin film spin-coated from 1:1 chloroform:acetonitrile at 20 °C) vapor doped with F$_4$TCNQ, as described in ref. [88].

### 3.4. Size of the Dopant

The size of a dopant can significantly affect the spatial separation between polarons located on the polymer backbone and the counterion, and thus the strength of the Coulomb interaction between the two species [Fig. 7(a)]. The distance between the $F_4TCNQ$ anion and a polaron on a P3HT backbone is thought to be about 5-9 Å,[35] which leads to a large energy barrier for dissociation and hence a low $\eta_{diss}$. Instead, large dopants such as dodecaborane (DDB) clusters with a diameter of 20 Å, which enforces a much larger distance between the polaron and counterion, result in highly mobile carriers.[54] DDB dopants of the form $B_{12}(OCH_3R)_{12}$ with R being a functionalized aryl group have an EA$_{dopant}$ that ranges from 4.9 to 5.7 eV, which yields p-dopants of different oxidation strengths.[58] DDB dopants enter crystalline domains of P3HT despite their size and yield highly mobile polarons, with $\eta_{ion}$ reaching 100 % for DDB-F$_{72}$ with the highest EA$_{dopant}$.[58] The consequence of highly mobile polarons is a conductivity that is significantly improved for a given polaron density compared to material doped with a smaller reference dopant such as $F_4TCNQ$. The importance to physically separate counterion and polaron was also supported by kMC simulations, which show that polaron-counterion interactions lower the overall conductivity compared to the hypothetical case where no counterions are present.[96, 98] On the other hand, large dopant molecules can induce structural changes in the polymer because they are nevertheless able to intercalate between the side chains in crystalline polymer domains.[54, 56] Further, at high dopant concentrations aggregates of dopant molecules tend to disrupt the nanostructure of the polymer, which usually deteriorates the (thermo)electric properties [see section 3.6]. Hence, it is important to distinguish between the influence of the dopant size on polaron-counterion interactions, and the concomitant impact on the nanostructure of the semiconductor. Thomas et al. have studied vapor-doping of PBTTT with NOPF$_6$ followed by exchange of the anion via a sequential solution-based process, which allowed to vary the counterion size from 5.2 to 8.2 Å without significantly changing the nanostructure of the polymer.[99] While the counterion size influenced polaron delocalization, the electrical conductivity was not affected by an increase in size, which indicates that the influence of other factors such as structural order and grain boundaries on charge transport are important to consider, as recently confirmed by Jacobs et al.[100]

Large dopants occupy more volume, which ultimately limits the polaron density [section 2.3]. The total DOS of a polymer is on the order of $10^{27}$ m$^{-3}$, i.e. the material can, at best, sustain one charge per cubic nanometer. P3HT, for example, has a density of 1.1 g cm$^{-3}$ and therefore features 4 thiophene rings per nm$^{-3}$. Each thiophene is, in principle, a redox site and if we assume a highest oxidation level of 33%, which is common for PEDOT,[101] then we obtain a charge-carrier density of about $10^{27}$ m$^{-3}$. If each charge is accompanied by a dopant that is, e.g., 1 nm in diameter then the highest obtainable charge-carrier density is about 5x10$^{26}$ m$^{-3}$, which has been reported for, e.g., P3HT doped with the molybdenum dithiolene complex Mo(tfd-COCF$_3$)$_3$,[102] which has a size of 11 to 14 Å.

Several studies have shown that incorporating small dopants such as FeCl$_3$ or F$_4$TCNQ into P3HT can lead to a decrease in the π-stacking distance.[34, 88, 90, 103] The origin of this reduction in the distance between adjacent polymer backbones upon doping is still under discussion. Some authors suggest geometrical arguments such as the intercalation of small molecules like F$_4$TCNQ between the alkyl side chains of P3HT.[86, 92-95] However, it was also reported that upon introduction of structurally similar small molecules, which do not dope the polymer, no change in the π-stacking distance occurs.[104, 105] This would imply that the doping process itself alters the crystalline packing motif. Liu et al. used density functional theory (DFT) to investigate this question and suggested that the delocalization of the polaron across multiple adjacent P3HT backbones leads to attractive forces that reduce the π-stacking distance.[106]

The size of dopant molecules also influences the ability to diffuse in the polymer. On the one hand, rapid diffusion is desirable for sequential doping of polymer films with dopant solutions or dopant vapor. However, with regard to long-term stability, diffusion of dopants is not desired because it will lead to gradual changes in dopant distribution. Neutral F$_4$TCNQ has a diffusion coefficient $D$ of about $10^{-11}$ cm$^2$ s$^{-1}$ in P3HT at 40 °C, which however decreases once the dopant has accepted an electron due to the electrostatic interaction between polarons and anions, i.e. the F$_4$TCNQ anion features a much lower $D \approx 10^{-13}$ cm$^2$ s$^{-1}$.[107] Hence, it is possible to sequentially dope polymer films, while the doped material becomes more stable due to a reduction in the diffusion coefficient [cf. section 5]. Diffusion of F$_4$TCNQ type dopants can be reduced further by replacing one of the cyano groups with, e.g., a methyl ester group,[107] which Moulé et al. have used to develop a highly sophisticated patterning technique.[108] F$_6$TCNNQ anions, which are larger than F$_4$TCNQ anions, display a smaller diffusion coefficient,[109] which can be beneficial for the stability of the doped state [see section 4].

Mixing small and large dopants might provide a way to exploit the advantages of dopants with various sizes, i.e. the ability of small dopants to diffuse more easily and the higher polaron-counterion distance facilitated by larger dopants. In this context, Liang et al. have shown that under certain conditions doping a P3HT film with a mixture of Mo(tfd)$_3$ and FeCl$_3$ can lead to a higher power factor than doping with either of the dopants alone.[56]

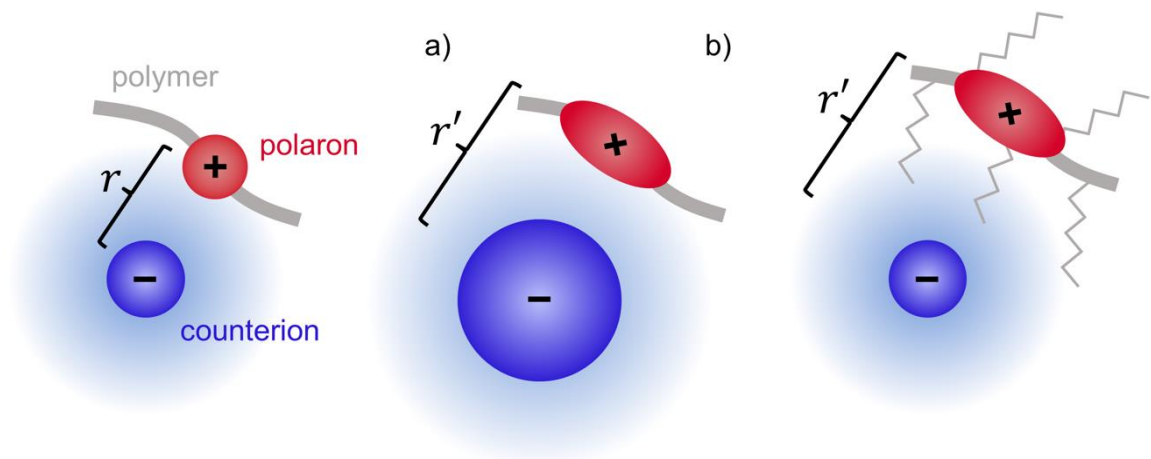

**FIG. 7**. (a) An increase in the size of the counterion or (b) suitable polymer side chains will increase the effective distance between a counterion and an adjacent polaron from $r$ to $r'$, resulting in a less strongly bound polaron due to a lower Coulomb interaction. As a result, the polaron will delocalize to a greater extent along the conjugated polymer backbone.

### 3.5. Spatial Distribution of Dopant Molecules

Charge transport strongly depends on the spatial distribution of the dopant. Clustering of dopants in the polymer can lead to a heavily-tailed DOS and hence a dramatic increase in the energetic disorder, which deteriorates the electrical conductivity.[110] Moreover, phase segregation between the polymer and the dopant can strongly limit $\eta_{ion}$ [see section 2.1]. One effective strategy that allows to control the miscibility of the two species and manipulate crystallization is the choice of suitable polymer side chains. Many dopants are polar molecules and hence conjugated polymers that feature more polar oligoether instead of alkyl side chains can increase the polymer-dopant miscibility.[111] Kroon et al. attached tetraethylene glycol side chains to a polythiophene backbone [cf. structure of p(g$_4$2T-T) in Fig. 2], which yields a conductivity of up to 100 S cm$^{-1}$ when doped with F$_4$TCNQ[111-113] or different acids.[23] Oligoether side chains increase the local dielectric constant, stabilize the dopant counterions and, as a result, significantly enhance the thermal stability [cf. section 4]. However, the use of polar side chains is not a universal solution to improve the thermal stability of the doped material. In addition, the IE$_{osc}$ of the polymer (in case of p-doping) as well as the precise interactions between dopant and the side chains must be taken into account.[114] Many molecular dopants prefer to intercalate between the side chains and hence a reduction in the side-chain grafting density increases the space available for the counterion and hence affects $\eta_{ion}$[115, 116] and the thermal stability.[117] The side-chain length is another parameter that influences polymer-dopant interactions as observed in case of PBTTT with different alkyl side chains ranging from *n*-octyl to *n*-octyldecyl.[118]

The spatial separation of the polymer backbone and dopant counterion strongly influences the extent of polaron delocalization since a more distant counterion results in a lower Coulomb binding energy [Fig. 7(b)].[119] Spano and co-workers have studied the infrared absorption spectra of P3HT doped with F$_4$TCNQ and correlated the oscillator strength and peak position of polaronic absorption bands with the extent of intra- and interchain polaron delocalization, which are influenced by factors such as the molecular weight and nanostructure of the polymer.[35] Again, the side chains can be used to modify the distance between the counterion and backbone [see section 3.4 for effect of dopant size]. For instance, Liu et al. studied DMBI-doped NDI-based copolymers with amphipathic side chains consisting of an alkyl spacer and an oligoether segment, which allowed to spatially separate counterions and the polymer backbone resulting in reduced Coulomb interactions.[119] In case of donor-acceptor copolymers, charge transfer only occurs if the dopant is located next to the right moiety, e.g. next to the donor moiety in case of p-doping, as shown by Di Nuzzo et al. for PCPDTBT doped with F$_4$TCNQ.[120]

### 3.6. Influence of Doping on the Nanostructure

Doping can strongly alter the nanostructure of a conjugated polymer in terms of the crystal structure, the degree of crystalline order, the crystal size and the connectivity between domains. The crystal structure of the semiconductor can change upon doping as exemplified by P3HT and PBTTT doped with F$_4$TCNQ, where the dopant intercalates between the side chains of the polymer.[89] Further, doping can enhance the degree of π-stacking of less ordered polymers such as p(g$_4$2T-T)[23] and regio-random P3HT.[105, 121] Likewise, doping of regio-regular P3HT can change the conjugation length of polymer chains in the amorphous regions, in particular of those that make up the rigid amorphous fraction, and hence improve connectivity and the overall conductivity.[122]

There are two principal doping methods: (1) co-processing of the polymer and dopant from the same solution and (2) sequential processing which involves casting of a polymer film followed by a doping step where the solidified film is brought in contact with a dopant solution or vapor [Fig. 8]. Co-processing allows to control the exact ratio between polymer and dopant. However, dissolving polymer and dopant in the same solvent can be challenging and the premature ionization of polymers in solution can reduce their solubility, leading to the formation of aggregates[86] that are characterized by less tie chains and hence a poor connectivity between crystalline domains, resulting in a lower charge-carrier mobility.[86] Hence, the processing solvent strongly influences the solid-state nanostructure of films obtained by, e.g., spin- or blade-coating. For instance, co-processing of P3HT with F$_4$TCNQ from chloroform results in a much larger degree of π-stacking and, hence $\eta_{doping}$ as compared to films co-processed from chlorobenzene in case of which pre-aggregation in solution was observed.[95]

Sequential processing requires that the dopant diffuses into the already solidified polymer.[123, 124] Hence, thin polymer films can be doped rapidly, while thicker structures must be exposed to a dopant solution or vapor for long periods of time since the diffusion coefficient of many dopants in the semiconductor host is relatively low [see section 3.4]. For instance, 50 nm thin films of P3HT can be sequentially doped with Mo(tfd-COCF$_3$)$_3$ within minutes,[102] while doping of tens of micrometers thick structures takes two days.[125] The two-step process ensures that the processing conditions that are used for preparation of the polymer film determine its texture and degree of in-plane orientation, which are largely preserved during the sequential doping step. For instance, P3HT films maintain an edge-on texture upon sequential processing, which benefits in-plane charge transport, while co-processing results in a more isotropic texture.[34, 86, 94] Further, thin films that feature long-range oriented domains[90, 126] or a high degree of in-plane uniaxial alignment[8, 102, 109] maintain their microstructure upon doping. Similar observations have been made for bulk materials such as foams of P3HT that keep their microstructure upon sequential doping with F$_4$TCNQ.[127] Sequential processing can be used to elucidate structure-property relationships relevant for charge transport in doped polymers. For instance, sequential doping of P3HT thin films with F$_4$TCNQ has shown that the crystallinity of the polymer strongly influences the charge-carrier mobility while $\eta_{ion}$ remains unaffected, overall resulting in an increase in $\sigma$ but not $\alpha$ with the degree of order.[87] Sequential processing tends to result in a higher thermoelectric power factor than co-processing,[90, 126] since the most optimal nanostructure can be selected prior to doping.

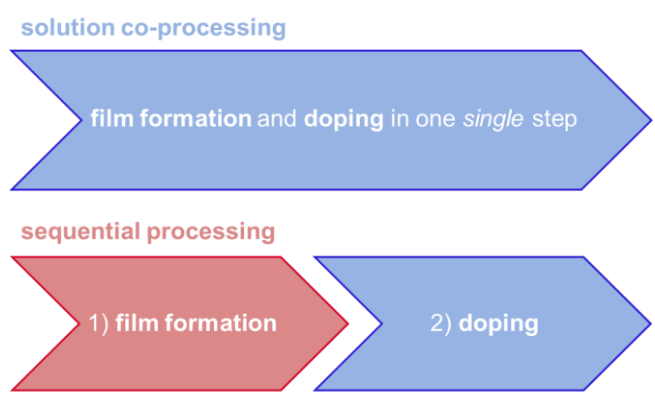

**FIG. 8.** Solution co-processing involves deposition of the organic semiconductor and molecular dopant from the same solution (top) while sequential processing is a two-step process where the conjugated polymer is first cast from solution followed by doping through contact with a molecular dopant dissolved in an orthogonal solvent or dopant vapor (bottom).

## 4. Stability of the Doped State

Thermoelectric energy harvesting requires prolonged exposure of p- and n-type materials to heat gradients, which necessitates a high degree of thermal stability. Many conjugated polymers display good chemical stability at ambient conditions but gradually undergo thermo-oxidative degradation at elevated temperatures above 100 °C, and certainly above 200 °C, unless stabilizers such as those commonly used for polyolefins are added.[128] Doped polymers with good thermal stability even at 100 °C are more difficult to realize. Hence, we argue that polymer based thermoelectric devices can only be used in combination with relatively cold heat sources whose temperature $T_{hot}$ is less than 200 °C, which ultimately limits the generated power since $P_{out} \propto \Delta T^2 = (T_{hot} - T_{cold})^2$.[129, 130]

The thermal stability strongly depends on the polymer:dopant pair. Some dopants are relatively small molecules with a high vapor pressure that can diffuse within a polymer,[107, 131] which can be exploited for sequential processing [cf. section 3.6], but also implies that dopants can escape the conducting material again [Fig. 9]. F$_4$TCNQ, for instance, can sublime above 100 °C, which means that a F$_4$TCNQ-doped polymer tends to gradually return to its neat, undoped state when heated.[40, 85] The polarity of the side chain can however improve the ability of the polymer to retain the dopant upon heating.[85, 132] Other dopants such as 1,3-propanedisulfonic acid (PDSA) have a lower vapor pressure and therefore p(g$_4$2T-T) p-doped with PDSA retains its high $\sigma$ of up to 120 S cm$^{-1}$ for at least 20 hours at 120 °C and exposure to air.[23]

Some dopants undergo side reactions with the polymer. F4TCNQ can abstract a hydrogen from P3HT –but not p(g$_4$2T-T)– resulting in the formation of a weak dopant, HF$_4$TCNQ$^-$, which strongly reduces the conductivity especially upon annealing at temperatures above 100 °C.[133] Further, F$_4$TCNQ initially undergoes integer charge transfer (ICT) with polymers such as P3HT and PBTTT but gradually reverts to a thermodynamically stable CTC between the polymer and the dopant.[134] The concomitant decrease in the number of charge carriers that can contribute to transport results in a gradual decrease in conductivity. The substrate temperature during vapor-doping of PBTTT with F$_4$TCNQ allows to adjust the degree of CTC formation.[40] Samples with a moderate amount of CTC states retain a higher conductivity upon prolonged annealing at 100 °C because the dopant is retained by the polymer to a greater extent, resulting in a higher thermal stability of the doped state.[40]

While many p-doped polymers are stable at ambient conditions, the electrical conductivity of many n-doped polymers rapidly diminishes upon exposure to air and water. Improved air stability is observed for polymers with an EA$_{osc}$ > 3.9 eV. While the electrical conductivity of PNDIT2 (EA$_{osc}$ = 3.8 eV) doped with N-DPBI decreases rapidly upon air exposure, a similar polymer with a lower EA$_{osc}$ = 4.0 eV, achieved by thionation of the naphthalene diimide segments, shows a steady $\sigma \approx 10^{-3}$ S cm$^{-1}$ for at least 1000 min at ambient conditions.[135] Thermally stable n-type materials have been reported for poly-(benzimidazobenzophenanthroline) (BBL; EA$_{osc}$ = 4.2 eV).[91] BBL, sequentially n-doped with TDAE or N-DMBI, initially features a relatively high $\sigma \approx 2$ S cm$^{-1}$, which rapidly drops in case of TDAE due to the low vapor pressure of the dopant while N-DMBI doped BBL displays a stable electrical conductivity at 190 °C for at least 20 hours under nitrogen.[136] Yang et al. have recently screened a series of triaminomethane based dopants in-silico and identified one candidate, TAM [Fig. 2(b)] that was subsequently synthesized, which yielded a stable $\sigma \approx 21$ S cm$^{-1}$ at ambient conditions, as well as $\alpha^2 \sigma \approx 51$ µW m$^{-1}$ K$^{-2}$, when used to n-dope the polymer FBDPPV [see Fig. 2 for chemical structure].[137]

Good thermal stability at 200 °C under nitrogen is also observed for all-polymer blends of BBL and p(g$_4$2T-T), which undergo ground-state electron transfer,[91] or BBL and poly(ethyleneimine) (PEI),[138] resulting in conducting materials that do not contain any small-molecular dopant, which could diffuse or sublime [Fig. 9]. All-polymer blends of BBL and PEI display an electron conductivity of up to $\sigma \approx 8$ S cm$^{-1}$, with micrometer-thick films also showing remarkable air-stability. Overall, these polymer blends can be considered an n-type equivalent to the most widely studied p-type conducting polymer material, PEDOT:PSS.

The poor air stability of some n-doped polymers can be mitigated through encapsulation. For instance, benzodifurandione based copolymers doped with N-DMBI show a stable electrical conductivity of up to 90 S cm$^{-1}$ once a 500 nm thick layer of an amorphous fluoropolymer is applied that effectively protects the n-doped polymer from air.[139] A protection layer may also allow to address the high vapor pressure of some dopants by keeping dopant molecules inside the material.

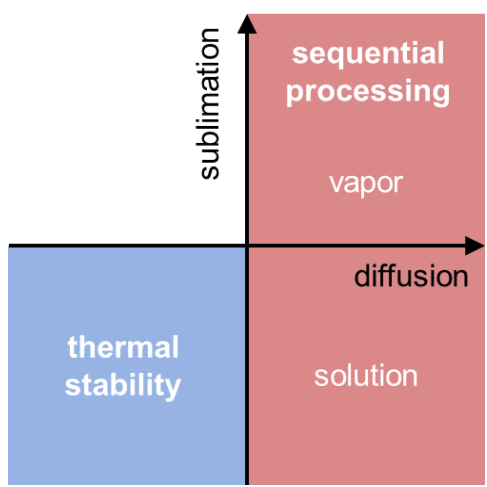

**FIG. 9.** Small-molecular dopants that easily sublime and display a high diffusion coefficient in the semiconductor are excellent for sequential processing (red) while larger (polymeric) dopants and/or those that strongly interact with the semiconductor may facilitate a higher degree of thermal stability (blue).

# 5. Effect of Doping on the Relationship Between Seebeck Coefficient and Electrical Conductivity

As discussed in the introduction, an important quantity to determine the figure of merit $zT$ of a thermoelectric material is the power factor and thus the functional relationship between $\alpha$ and $\sigma$. While the general trends are well established, the details are less clear and significant deviations from the $\alpha \propto \sigma^{-1/4}$ proportionality proposed by Glaudell et al. have been reported, especially at higher doping levels. Following the discussion in section 3.2, this could at least partly arise due to dopant-induced changes to the DOS.[110, 140] Doping can also alter the nano- and microstructure of the semiconductor [cf. sections 3.5 and 3.6]. In this context, Thomas et al. have studied electrochemical oxidation of PBTTT and found that the exponent of the $\alpha$ vs. $\sigma$ relation can deviate from $m = -1/4$ [see section 1] even though the nanostructure was not altered.[141] Instead, the authors argue that doping induced energetic disorder results in a broadening of the local DOS. An alternative strategy to modify the DOS distribution is blending of two semiconductors A and B with different energy levels.[142, 143] This leads to a binary mixture A:B, exhibiting a double-peaked DOS. For an optimized mixture, the DOS can be tuned in a way that the transport energy $E_{\text{tr}}$ is shifted into the B-part of the DOS, while the Fermi energy remains in the A-part. This leads to high thermopowers at reasonable conductivities and moreover a power factor of the mixture that can exceed that of the neat compounds.[143]

It is also possible to manipulate the $\alpha$ vs. $\sigma$ relation and with this the power factor of doped polymers using sequential processing methods that introduce uniaxial alignment in polymer films, e.g. through high-temperature rubbing or drawing, followed by doping. While several studies have shown that the resulting alignment of the polymer backbone leads to a considerable increase of the electrical conductivity along the direction of orientation [see section 3], the influence on the thermopower is less clear. Sarabia-Riquelme et. al investigated the thermoelectric properties of wet-spun PEDOT:PSS fibers and found an increase in electrical conductivity by drawing the fibers in a DMSO bath.[144] This increase in conductivity was attributed to a synergistic effect of a larger degree of $\pi$-stacking of PEDOT upon removal of PSS through treatment with DMSO as well as the alignment of PEDOT upon drawing. Meanwhile, $\alpha$ was unaffected along the direction of orientation. Similar observations were also made by other authors for PEDOT:PSS based films and fibers.[145, 146] In contrast, several recent reports on PBTTT based thin films indicate that both $\alpha$ and $\sigma$ can be enhanced simultaneously,[8, 109, 118] resulting in a drastic improvement of the power factor by several orders of magnitude along the direction of orientation [Fig. 10].[8] The $\alpha$ vs. $\sigma$ relation has a universal shape for values measured both parallel and perpendicular to the alignment direction that, for PBTTT, is independent of the type of dopant (F$_4$TCNQ, F$_6$TCNNQ or FeCl$_3$).[109] However, the side-chain length influences the degree of anisotropy in thermoelectric parameters that can be obtained.[147] For films of P3HT doped with F$_4$TCNQ it was recently shown that the degree of anisotropy in $\sigma$ and $\alpha$ depends on the initial structure of the film and shows a discontinuity with the transition from a smetic-like to a semicrystalline phase.[148] In this context, it seems reasonable to expect that the experimental conditions used to align P3HT determine whether an anisotropy in $\alpha$ is observed[8, 89, 102] or not.[125] The assumption that the value of $\alpha$ depends on the packing motif of the polymer is supported by kMC simulations,[147] which indicate that different lattice structures affect the degree of anisotropy in $\sigma$ and $\alpha$, emphasizing the importance of the nanostructure on the thermoelectric properties.

Several recent studies have observed that for both initially n- as well as p-doped polymers a sign change in the Seebeck coefficient occurs at high dopant concentrations. Hwang et al. experienced a sign change in the Seebeck coefficient in poly(pyridinium phenylene).[149] The authors argued that, as a result of extensive doping, the original LUMO level is fully filled and thus acts as a new HOMO level. However, this requires high dopant concentrations. In contrast, it was shown that both NDI[140] and DPP[150] based polymers, which are n-doped at low dopant concentrations, already show such a change in the Seebeck coefficient at much lower dopant concentrations. In case of the polymer PNDI2TEG-2T this behavior was explained by the formation of new mobile states upon doping that lie below the LUMO of the undoped material. Simulations suggested that the transport energy can cross the Fermi energy even at moderate dopant densities, leading to a sign change in $\alpha$ as long as the formed CTC states show a higher localization length than the host states, i.e. the CTC states are more delocalized.[140] Interestingly, in case of both polymers, PNDI2TEG-2T and PNDI2TEG-2Tz, the sign change in $\alpha$ disappeared upon modification of the donor moiety, which thus provides a means to tailor the DOS.[140]

Likewise, the Seebeck coefficient of several p-doped polymers shows a sign change from positive to negative upon an increase in concentration of the oxidizing agents FeCl$_3$ or NOBF$_4$.[65, 151] For non-oriented PDPP-5T thin films the doping concentration above which a change in the sign of $\alpha$ occurs increases with molecular weight. The authors attributed the sign change to state filling in a purely Gaussian DOS and modification of the width of the latter with molecular weight.[151] However, the effect of Coulomb interactions on the shape of the DOS [see section 3.2] is ignored and, moreover, the assumed DOS is not in agreement with the ones measured in refs. [140] and 65.

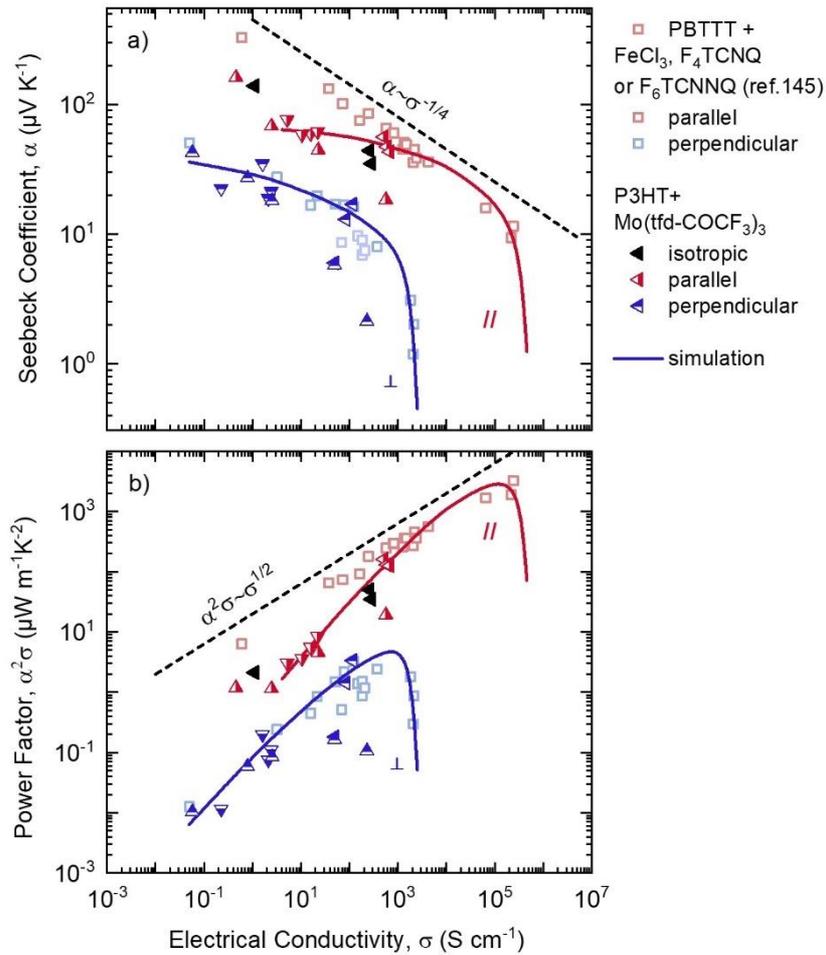

**FIG. 10. Impact of uniaxial orientation on the thermoelectric parameters:** (a) Seebeck coefficient $\alpha$ and (b) power factor $\alpha^2\sigma$ as a function of electrical conductivity $\sigma$ for P3HT (triangles) and PBTTT (squares) measured parallel (red) and perpendicular to the alignment direction (blue); black data points are values for isotropic P3HT samples. Full lines correspond to simulations based on a kMC model applied to a regular lattice, with an anisotropy ratio of $\xi_\parallel/\xi_\perp = 4$, where $\xi_\parallel$ and $\xi_\perp$ are the localization length parallel and perpendicular to the alignment direction, respectively (for details see ref. [147]). Experimental values are taken from ref. [147] (□), ref. [102] (◄), ref. [89] (▼) and ref. [8] (▲). Note that this picture is a modified version of Figure 4 in ref. [147] (D. Scheunemann, V. Vijayakumar, H. Zeng, P. Durand, N. Leclerc, M. Brinkmann, M. Kemerink, Adv. Electron. Mater., 6, 2000218, 2020; licensed under a Creative Commons Attribution (CC BY 4.0) license.), including additional data for P3HT.

## 6. Interplay Between Electrical and Thermal Conductivity

The thermal conductivity $\kappa$ is given by the sum of the electronic contribution $\kappa_{el}$ and the phonon contribution $\kappa_{ph}$:[152]

$$\kappa = \kappa_{el} + \kappa_{ph} \tag{6.1}$$

Since free electrons carry both heat and charge, both $\sigma$ and $\kappa$ depend on the charge-carrier concentration and are hence related. The Wiedemann-Franz (WF) law,[153] which is valid for many metals and metal alloys where electrons only experience elastic scattering, states that the ratio $\kappa_{el}/\sigma$ is approximately constant:

$$\frac{\kappa_{el}}{\sigma} = L_0 T \tag{6.2}$$

where $L_0 = \left(\frac{\pi^2}{3}\right)\left(\frac{k_B}{e}\right)^2 = 2.44 \times 10^{-8}\, W\,\Omega\,K^{-2}$ is the Sommerfeld value of the Lorenz number. Attempts to apply the WF law to disordered organic materials has met with mixed results[154-156] since electronic charge transport is often dominated by inelastic hopping between redox sites. For instance, Weathers et al. have observed for suspended films of PEDOT:PSS and PEDOT:Tos that the increase of the in-plane component of $\kappa$ with $\sigma$ exceeds the trend predicted by the WF law for metals, which indicates significant electronic thermal transport.[155]

To examine the ratio $\kappa_{el}/\sigma$ in organic systems, starting from the (generalized) Miller and Abrahams hopping transfer rate,[157] different theories have recently been proposed, leading to the definition of new Lorenz numbers.

Assuming a system of identical molecular charge transfer sites, Craven et al. have derived a new Molecular-Wiedemann-Franz (MWF) law for amorphous molecular solids.[158] The MWF law states that:

$$\frac{\kappa_{el}}{\sigma} = L_M T_M \tag{6.3}$$

where $L_M = (k_B/e)^2$ is the molecular Lorenz number, and $T_M = \lambda/k_B$ is an effective temperature parameterized by the reorganization energy $\lambda$ of the redox charge transfer. While the WF law is material-independent, the new MWF law implicitly depends on the temperature through $\lambda$. Since $\lambda$ often depends only weakly on temperature, the MWF will be temperature-independent in most cases.

Scheunemann et al.[159] have extended the transport models for $\alpha$ and $\sigma$ developed by Schmechel et al.[160] and Ihnatsenka et al.[161] to $\kappa_{el}$. The extended model shows that the energetic disorder strongly influences the ratio $\kappa_{el}/\sigma$. In particular, $\kappa_{el}/\sigma$ is found to be smaller than the Sommerfeld value in weakly disordered systems with high charge-carrier concentrations [see Fig. 11].

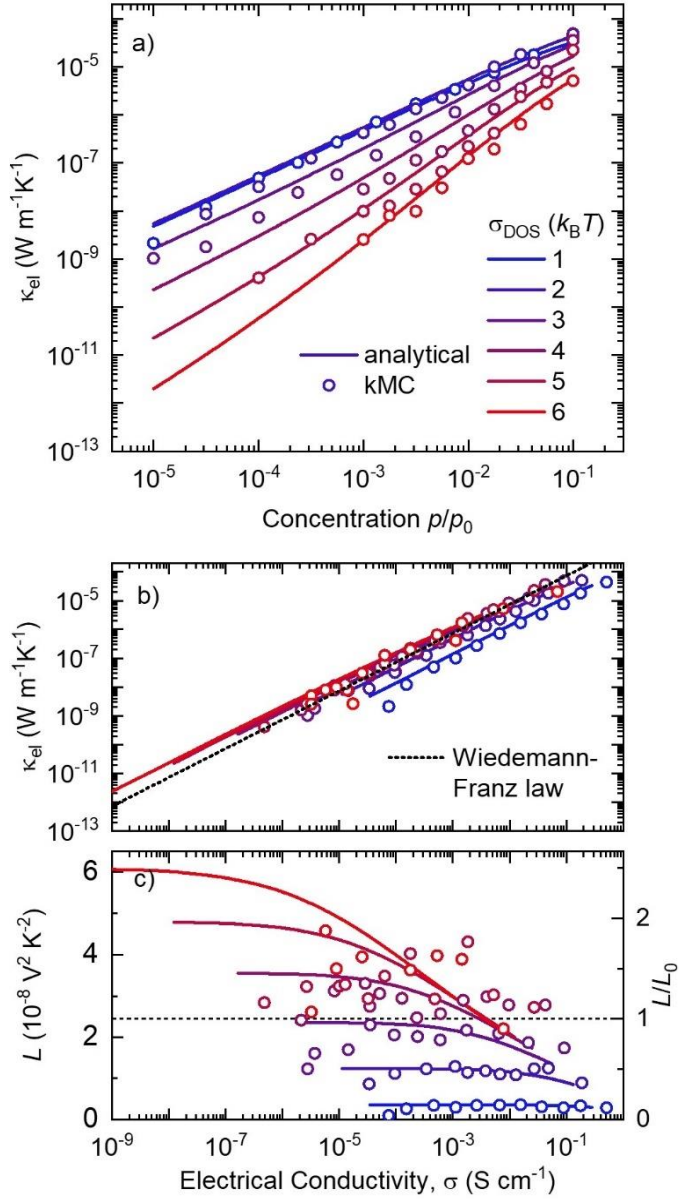

**FIG. 11. Influence of doping on the thermal conductivity:** $\kappa_{el}$ from the analytical model (solid lines) and kMC simulations (symbols) for different energetic disorder $\sigma_{DOS}$ with dependency on (a) relative carrier concentration $p/p_0$ and (b) electrical conductivity. The dotted black line in (b) represents the result of the WF law with $L = L_0$. (c) Lorenz number $L$ as a function of $\sigma$. The dashed line indicates the Sommerfeld value for a free electron gas.[159] Figure reproduced from ref. [159] (D. Scheunemann, M. Kemerink, Phys. Rev. B, 101, 075206, 2020; licensed under a Creative Commons Attribution (CC BY 4.0) license.).

The validity of the WF law in thermoelectric materials is of great interest as the ratio $\kappa_{el}/\sigma$ naturally appears in the figure of merit $zT$. In fact, combining equations 1.1, 6.1 and 6.2 yields:

$$zT = \frac{\alpha^2 \sigma T}{\kappa_{ph} + \sigma \mathcal{L}\tau} \tag{6.4}$$

where $\mathcal{L}$ is either $L_0$ or a non-Sommerfeld Lorenz number, which potentially is a function of the charge-carrier concentration and of the energetic disorder,[159] and $\tau$ is the absolute temperature $T$ or an effective temperature $T_M$, depending on the model that better describes the ratio $\kappa_{el}/\sigma$. On the one hand, if $k_{ph} \gg \sigma\mathcal{L}\tau$, then $zT \propto \alpha^2\sigma$ and one can enhance $zT$ by finding an optimal charge-carrier concentration that maximizes the power factor $\alpha^2\sigma$, assuming that doping does not affect $\kappa_{ph}$. The latter may not apply in the case of molecular doping, which can alter the nanostructure and hence affect charge transport [section 3.6]. Interestingly, for PBTTT doped with $F_4$TCNQ, Zapata-Arteaga et. al. have shown that phonon scattering through the presence of dopant molecules can reduce $\kappa$ from about 0.7 to 0.4 W m$^{-1}$ K$^{-1}$.[162] On the other hand, if $k_{ph} \ll \sigma\mathcal{L}\tau$, then $zT \sim \alpha^2/\mathcal{L}$ and one can enhance $zT$ by maximizing the ratio $\alpha^2/\mathcal{L}$. The crucial parameter here is likely the Seebeck coefficient, since the square of $\alpha$ probably varies more than the Lorentz factor in most cases. In ref. [159], however, it was shown that in both cases the energetic disorder of the material is decisive for high $zT$ values.

Recent studies show that $\kappa_{el}$ becomes the dominant quantity of $\kappa$ only at sufficiently high $\sigma$,[159, 162] and that $\kappa_{ph}$ should be reduced below 0.2 W m$^{-1}$ K$^{-1}$ while maintaining a large $\alpha$ in order to achieve a high $zT$.[129] In this context, Liu et al. have recently demonstrated that some fullerene derivatives, which are generally known to display a very low $\kappa_{ph} < 0.1$,[163] can maintain a high $\alpha$ of about -250 µV K$^{-1}$ upon doping with N-DMBI, resulting in a figure of merit of up to $zT$ = 0.34 at 120 °C.[164]

Fig. 12 shows iso-$zT$ lines at 300 K calculated according to the original WF law, to the model from Scheunemann et al., and to the model of Craven et al. (for two different values of $\lambda$). The figure shows that, depending on the model, $zT$ benefits differently from increasing $\sigma$, and that the iso-$zT$ lines calculated with the model of Craven et al. tend to overlap with the iso-$zT$ lines obtained from the model of Scheunemann et al. for $\lambda \approx 0.05$ eV.

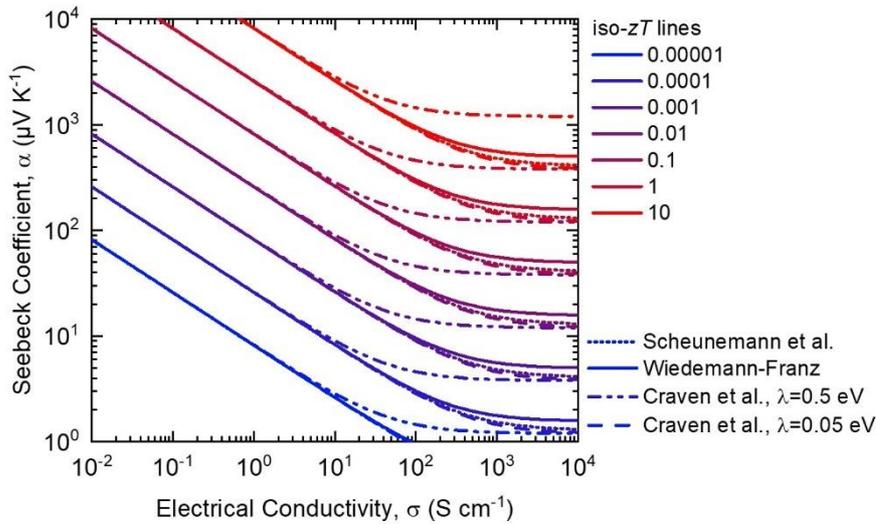

**FIG. 12. Interplay of thermoelectric parameters for different models for κ$_{el}$:** Iso-$zT$ lines in the $\sigma - \alpha$ plane calculated according to the original WF law (continuous lines), the model of Craven et al. assuming $\lambda = 0.5\ eV$ and $\lambda = 0.05\ eV$ (dashed and dot-dashed lines, respectively), and the model from ref. [159] (dotted lines) assuming $k_{ph}$ = 0.2 W m$^{-1}$ K$^{-1}$.

Organic materials can display anisotropic charge transport that depends on the nano- and microstructure [section 3.6]. Both measurements[154, 165] and kMC simulations[147] have demonstrated that such a structural anisotropy can also lead to anisotropic $\kappa_{el}$. Assessment of the validity of the WF law for such anisotropic systems is not trivial as the measurement of $\kappa$ and $\sigma$ must be done along the same direction. The development of experimental techniques that permit to measure $\kappa$ and $\sigma$ along the same direction or, preferably, along all non-equivalent directions is therefore a prerequisite for gaining detailed knowledge about the relation between $\sigma$ and $\kappa_{el}$. Moreover, structural anisotropy can be accompanied by significant differences in $\kappa_{ph}$ in the parallel and perpendicular direction of orientation, which is reasonable when considering that changes in the nanostructure can yield a high $\kappa_{ph}$ of above 1 W m$^{-1}$ K$^{-1}$ also for conjugated polymers.[166]

## 7. Outlook

We have discussed the interplay of the semiconductor nanostructure, processing, doping efficiency and charge transport, which is increasingly well understood in qualitative terms. More work is needed to enable a complete quantitative description of the electrical properties of doped organic semiconductors, and here the dialogue between experimentalists and theorists will continue to be instrumental. For example, the impact of alignment of the semiconductor in thin-film or bulk geometries is emerging as a powerful means that allows to not only maximize the electrical conductivity but also to constrain the complexity of doped materials, which facilitates new theoretical understanding. Rationales that aid the selection of dopants for a given organic semiconductor, and vice versa, with the aim to optimize the processability, doping efficiency and stability are slowly emerging. It can be anticipated that further synthetic work, carried out in vitro and increasingly in silico, will unearth new design principles.

A compromise will likely need to be struck with regard to the dopant size. Sufficiently small dopant molecules facilitate processing concepts such as vapor doping and enable sequential processing (doping) of thicker structures, since small molecules tend to diffuse more rapidly. On the other hand, larger molecules diffuse slowly, or not at all, which benefits the long-term stability of the doped state. In this context, the use of all-polymer systems, comprising both conjugated polymers doped or complexed with polymeric dopants or polyelectrolytes (cf. PEDOT:PSS and BBL:PEI) as well as pairs of conjugated polymers that undergo ground-state electron transfer may facilitate a new generation of stable organic conductors. All-polymeric materials may allow to circumvent the high reactivity of many redox dopants and acids, which raises both environmental concerns and prevents their use for applications in bioelectronics and wearable electronics, where benign materials are often a prerequisite.

Another aspect that we expect to grow in importance is the maximum number of charges that can be generated per given volume of material, which is limited not only by the doping efficiency but also the size of dopant molecules. Larger dopants tend to benefit charge dissociation but, on the other hand, take up a larger volume that cannot be occupied by the ionized semiconductor where charge transport ultimately occurs.

With respect to thermoelectric applications, both experimental observations and theoretical predictions have shown that fertile new avenues such as alignment of the polymer backbone and raising of the dielectric constant of a polymer benefit both the thermopower and electrical conductivity. One important question, however, is whether the resulting increase in the power factor will be accompanied by changes in the thermal conductivity. For instance, alignment of the polymer backbone can be expected to significantly increase the lattice component of the thermal conductivity. The design of new high-performance thermoelectric materials must therefore pay more attention to the still poorly understood thermal conductivity of doped polymers. In this respect, the field should continue to harness the highly productive synergies that arise through the combination of experiments and computational approaches.

Finally, we would like to point out that little attention is currently being paid to the environmental burden involved in processing from organic solvents as well as recycling of doped organic materials – and organic (semi)conductors in general– at the end of their lifetime. A complete life cycle assessment will be needed to ensure that organic thermoelectrics evolves into a sustainable technology. We argue that it would be advantageous if all types of materials that are needed to fabricate a wide range of organic electronic devices, from p- and n-type semiconductors and conductors to electrolytes and insulators, can be realized with only a handful of different materials that can, ideally, be separated again into their individual components. Ultimately, this vision may allow to develop green organic electronics that help to significantly reduce electronic waste.

## Data Availability

The data that support the findings of this study are available from the corresponding author upon reasonable request.

## Acknowledgements

We gratefully acknowledge financial support from the Swedish Research Council (grants no. 2016-06146 and 2018-03824), the European Union's Horizon 2020 research and innovation programme under the Marie Skłodowska-Curie grant agreement No. 799477 (HyThermEl) and No. 955837 (HORATES) and the Knut and Alice Wallenberg Foundation through a Wallenberg Academy Fellowship Prolongation grant. We thank Jonna Hynynen for experimental contributions and Renee Kroon for synthesizing p(g$_4$2T-T).

## References

[1]T. J. Seebeck, "Ueber die magnetische Polarisation der Metalle und Erze durch Temperaturdifferenz," Annalen der Physik **82**, 253-286 (1826).

[2]H. Fritzsche, "A general expression for the thermoelectric power," Solid State Commun. **9**, 1813-1815 (1971).

[3]A. M. Glaudell, J. E. Cochran, S. N. Patel, and M. L. Chabinyc, "Impact of the doping method on conductivity and thermopower in semiconducting polythiophenes," Adv. Energy Mater. **5**, 1401072 (2015).

[4]D. Beretta, N. Neophytou, J. M. Hodges, M. G. Kanatzidis, D. Narducci, M. Martin- Gonzalez, M. Beekman, B. Balke, G. Cerretti, W. Tremel, A. Zevalkink, A. I. Hofmann, C. Müller, B. Dörling, M.

Campoy-Quiles, and M. Caironi, "Thermoelectrics: From history, a window to the future," Mater. Sci. Eng. R Rep. **138**, 100501 (2019).

[5] G. Zuo, H. Abdalla, and M. Kemerink, "Impact of doping on the density of states and the mobility in organic semiconductors," Phys. Rev. B **93**, 235203 (2016).

[6] A. J. Heeger, "Semiconducting and metallic polymers: The fourth generation of polymeric materials," J. Phys. Chem. B **105**, 8475-8491 (2001).

[7] K. Akagi, H. Shirakawa, K. Araya, A. Mukoh, and T. Narahara, "Highly conducting polyacetylene films prepared in a liquid crystal solvent," Polym. J **19**, 185-189 (1987).

[8] V. Vijayakumar, Y. Zhong, V. Untilova, M. Bahri, L. Herrmann, L. Biniek, N. Leclerc, and M. Brinkmann, "Bringing conducting polymers to high order: Toward conductivities beyond $10^5$ S cm$^{-1}$ and thermoelectric power factors of 2 mW m$^{-1}$ K$^{-2}$," Adv. Energy Mater. **9**, 1900266 (2019).

[9] A. Elschner, S. Kirchmeyer, W. Lovenich, U. Merker, and K. Reuter, *PEDOT: Principles and applications of an intrinsically conductive polymer*, CRC Press, 2010.

[10] O. Bubnova, Z. U. Khan, A. Malti, S. Braun, M. Fahlman, M. Berggren, and X. Crispin, "Optimization of the thermoelectric figure of merit in the conducting polymer poly(3,4-ethylenedioxythiophene)," Nat. Mater **10**, 429-433 (2011).

[11] G. H. Kim, L. Shao, K. Zhang, and K. P. Pipe, "Engineered doping of organic semiconductors for enhanced thermoelectric efficiency," Nat. Mater **12**, 719-723 (2013).

[12] D. A. Mengistie, C.-H. Chen, K. M. Boopathi, F. W. Pranoto, L.-J. Li, and C.-W. Chu, "Enhanced thermoelectric performance of PEDOT:PSS flexible bulky papers by treatment with secondary dopants," ACS Applied Materials & Interfaces **7**, 94-100 (2015).

[13] O. Bubnova, and X. Crispin, "Towards polymer-based organic thermoelectric generators," Energy Environ. Sci. **5**, 9345-9362 (2012).

[14] Z. Fan, and J. Ouyang, "Thermoelectric properties of PEDOT:PSS," Adv. Electron. Mater. **5**, 1800769 (2019).

[15] M. Massetti, F. Jiao, A. J. Ferguson, D. Zhao, K. Wijeratne, A. Würger, J. L. Blackburn, X. Crispin, and S. Fabiano, "Unconventional thermoelectric materials for energy harvesting and sensing applications," Chemical Reviews **121**, 12465-12547 (2021).

[16] J.-C. Chiang, and A. G. MacDiarmid, "'Polyaniline': Protonic acid doping of the emeraldine form to the metallic regime," Synth. Met. **13**, 193-205 (1986).

[17] E. M. Thomas, E. C. Davidson, R. Katsumata, R. A. Segalman, and M. L. Chabinyc, "Branched side chains govern counterion position and doping mechanism in conjugated polythiophenes," ACS Macro Lett. **7**, 1492-1497 (2018).

[18]B. D. Naab, S. Guo, S. Olthof, E. G. B. Evans, P. Wei, G. L. Millhauser, A. Kahn, S. Barlow, S. R. Marder, and Z. Bao, "Mechanistic study on the solution-phase n-doping of 1,3-Dimethyl-2-aryl-2,3-dihydro-1H-benzoimidazole derivatives," J. Am. Chem. Soc. **135**, 15018-15025 (2013).

[19]S. Jhulki, H.-I. Un, Y.-F. Ding, C. Risko, S. K. Mohapatra, J. Pei, S. Barlow, and S. R. Marder, "Reactivity of an air-stable dihydrobenzoimidazole n-dopant with organic semiconductor molecules," Chem **7**, 1050-1065 (2021).

[20]R. A. Schlitz, F. G. Brunetti, A. M. Glaudell, P. L. Miller, M. A. Brady, C. J. Takacs, C. J. Hawker, and M. L. Chabinyc, "Solubility-limited extrinsic n-type doping of a high electron mobility polymer for thermoelectric applications," Adv. Mater. **26**, 2825-2830 (2014).

[21]S. Riera-Galindo, A. Orbelli Biroli, A. Forni, Y. Puttisong, F. Tessore, M. Pizzotti, E. Pavlopoulou, E. Solano, S. Wang, G. Wang, T.-P. Ruoko, W. M. Chen, M. Kemerink, M. Berggren, G. di Carlo, and S. Fabiano, "Impact of singly occupied molecular orbital energy on the n-doping efficiency of benzimidazole derivatives," ACS Appl. Mater. Interfaces **11**, 37981-37990 (2019).

[22]H. Guo, C.-Y. Yang, X. Zhang, A. Motta, K. Feng, Y. Xia, Y. Shi, Z. Wu, K. Yang, J. Chen, Q. Liao, Y. Tang, H. Sun, H. Y. Woo, S. Fabiano, A. Facchetti, and X. Guo, "Transition metal-catalysed molecular n-doping of organic semiconductors," Nat. Mater. **599**, 67-73 (2021).

[23]A. I. Hofmann, R. Kroon, L. Yu, and C. Müller, "Highly stable doping of a polar polythiophene through co-processing with sulfonic acids and bistriflimide," J. Mater. Chem. C **6**, 6905-6910 (2018).

[24]B. Yurash, D. X. Cao, V. V. Brus, D. Leifert, M. Wang, A. Dixon, M. Seifrid, A. E. Mansour, D. Lungwitz, T. Liu, P. J. Santiago, K. R. Graham, N. Koch, G. C. Bazan, and T.-Q. Nguyen, "Towards understanding the doping mechanism of organic semiconductors by Lewis acids," Nat. Mater. **18**, 1327-1334 (2019).

[25]M. Arvind, C. E. Tait, M. Guerrini, J. Krumland, A. M. Valencia, C. Cocchi, A. E. Mansour, N. Koch, S. Barlow, S. R. Marder, J. Behrends, and D. Neher, "Quantitative analysis of doping-induced polarons and charge-transfer complexes of poly(3-hexylthiophene) in solution," J. Phys. Chem. B **124**, 7694-7708 (2020).

[26]A. D. Scaccabarozzi, A. Basu, F. Aniés, J. Liu, O. Zapata-Arteaga, R. Warren, Y. Firdaus, M. I. Nugraha, Y. Lin, M. Campoy-Quiles, N. Koch, C. Müller, L. Tsetseris, M. Heeney, and T. D. Anthopoulos, "Doping approaches for organic semiconductors," Chem. Rev. **122**, 4420-4492 (2022).

[27]W. Zhao, J. Ding, Y. Zou, C.-a. Di, and D. Zhu, "Chemical doping of organic semiconductors for thermoelectric applications," Chem. Soc. Rev. **49**, 7210-7228 (2020).

[28]I. Salzmann, G. Heimel, M. Oehzelt, S. Winkler, and N. Koch, "Molecular electrical doping of organic semiconductors: fundamental mechanisms and emerging dopant design rules," Acc. Chem. Res. **49**, 370-378 (2016).

[29] I. E. Jacobs, and A. J. Moulé, "Controlling molecular doping in organic semiconductors," Adv. Mater. **29**, 1703063 (2017).

[30] T. Biskup, "Doping of organic semiconductors: Insights from EPR spectroscopy," Applied Physics Letters **119**, (2021).

[31] A. I. Hofmann, R. Kroon, S. Zokaei, E. Järsvall, C. Malacrida, S. Ludwigs, T. Biskup, and C. Müller, "Chemical doping of conjugated polymers with the strong oxidant magic blue," Adv. Electron. Mater. **6**, 2000249 (2020).

[32] I. Sahalianov, J. Hynynen, S. Barlow, S. R. Marder, C. Müller, and I. Zozoulenko, "UV-to-IR Absorption of Molecularly p-Doped Polythiophenes with Alkyl and Oligoether Side Chains: Experiment and Interpretation Based on Density Functional Theory," J. Phys. Chem. B **124**, 11280-11293 (2020).

[33] T. L. Murrey, M. A. Riley, G. Gonel, D. D. Antonio, L. Filardi, N. Shevchenko, M. Mascal, and A. J. Moule, "Anion Exchange Doping: Tuning Equilibrium to Increase Doping Efficiency in Semiconducting Polymers," J. Phys. Chem. Lett. **12**, 1284-1289 (2021).

[34] D. T. Scholes, P. Y. Yee, J. R. Lindemuth, H. Kang, J. Onorato, R. Ghosh, C. K. Luscombe, F. C. Spano, S. H. Tolbert, and B. J. Schwartz, "The effects of crystallinity on charge transport and the structure of sequentially processed $F_4TCNQ$-doped conjugated polymer films," Adv. Funct. Mater. **27**, 1702654 (2017).

[35] R. Ghosh, A. R. Chew, J. Onorato, V. Pakhnyuk, C. K. Luscombe, A. Salleo, and F. C. Spano, "Spectral signatures and spatial coherence of bound and unbound polarons in P3HT films: theory versus experiment," J. Phys. Chem. C **122**, 18048-18060 (2018).

[36] J. Li, I. Duchemin, O. M. Roscioni, P. Friederich, M. Anderson, E. Da Como, G. Kociok-Köhn, W. Wenzel, C. Zannoni, D. Beljonne, X. Blase, and G. D'Avino, "Host dependence of the electron affinity of molecular dopants," Mater. Horiz. **6**, 107-114 (2019).

[37] B. Nell, K. Ortstein, O. V. Boltalina, and K. Vandewal, "Influence of dopant–host energy level offset on thermoelectric properties of doped organic semiconductors," J. Phys. Chem. C **122**, 11730-11735 (2018).

[38] A. Fediai, A. Emering, F. Symalla, and W. Wenzel, "Disorder-driven doping activation in organic semiconductors," Phys. Chem. Chem. Phys. **22**, 10256-10264 (2020).

[39] I. E. Jacobs, C. Cendra, T. F. Harrelson, Z. I. Bedolla Valdez, R. Faller, A. Salleo, and A. J. Moulé, "Polymorphism controls the degree of charge transfer in a molecularly doped semiconducting polymer," Mater. Horiz. **5**, 655-660 (2018).

[40] O. Zapata-Arteaga, B. Dörling, A. Perevedentsev, J. Martín, J. S. Reparaz, and M. Campoy-Quiles, "Closing the stability–performance gap in organic thermoelectrics by adjusting the partial to integer charge transfer ratio," Macromolecules **53**, 609-620 (2020).

[41] H. Méndez, G. Heimel, A. Opitz, K. Sauer, P. Barkowski, M. Oehzelt, J. Soeda, T. Okamoto, J. Takeya, J.-B. Arlin, J.-Y. Balandier, Y. Geerts, N. Koch, and I. Salzmann, "Doping of organic semiconductors: Impact of dopant strength and electronic coupling," Angew. Chem. Int. Ed. **52**, 7751-7755 (2013).

[42] H. Méndez, G. Heimel, S. Winkler, J. Frisch, A. Opitz, K. Sauer, B. Wegner, M. Oehzelt, C. Röthel, S. Duhm, D. Többens, N. Koch, and I. Salzmann, "Charge-transfer crystallites as molecular electrical dopants," Nat. Commun. **6**, 8560 (2015).

[43] M. Goel, M. Siegert, G. Krauss, J. Mohanraj, A. Hochgesang, D. C. Heinrich, M. Fried, J. Pflaum, and M. Thelakkat, "HOMO–HOMO electron transfer: An elegant strategy for p-type doping of polymer semiconductors toward thermoelectric applications," Adv. Mater. **32**, 2003596 (2020).

[44] G. Krauss, A. Hochgesang, J. Mohanraj, and M. Thelakkat, "Highly Efficient Doping of Conjugated Polymers Using Multielectron Acceptor Salts," Macromol. Rapid Commun. **42**, 2100443 (2021).

[45] D. Kiefer, R. Kroon, A. I. Hofmann, H. Sun, X. Liu, A. Giovannitti, D. Stegerer, A. Cano, J. Hynynen, L. Yu, Y. Zhang, D. Nai, T. F. Harrelson, M. Sommer, A. J. Moulé, M. Kemerink, S. R. Marder, I. McCulloch, M. Fahlman, S. Fabiano, and C. Müller, "Double doping of conjugated polymers with monomer molecular dopants," Nat. Mater. **18**, 149-155 (2019).

[46] H. T. Yi, N. Gartstein, and V. Podzorov, "Charge carrier coherence and Hall effect in organic semiconductors," Sci. Rep. **6**, 23650 (2016).

[47] P. Pingel, and D. Neher, "Comprehensive picture of p-type doping of P3HT with the molecular acceptor $F_4$TCNQ," Phys. Rev. B **87**, 115209 (2013).

[48] M. Schwarze, C. Gaul, R. Scholz, F. Bussolotti, A. Hofacker, K. S. Schellhammer, B. Nell, B. D. Naab, Z. Bao, D. Spoltore, K. Vandewal, J. Widmer, S. Kera, N. Ueno, F. Ortmann, and K. Leo, "Molecular parameters responsible for thermally activated transport in doped organic semiconductors," Nat. Mater. **18**, 242-248 (2019).

[49] A. Mityashin, Y. Olivier, T. Van Regemorter, C. Rolin, S. Verlaak, N. G. Martinelli, D. Beljonne, J. Cornil, J. Genoe, and P. Heremans, "Unraveling the mechanism of molecular doping in organic semiconductors," Adv. Mater. **24**, 1535-1539 (2012).

[50] M. L. Tietze, J. Benduhn, P. Pahner, B. Nell, M. Schwarze, H. Kleemann, M. Krammer, K. Zojer, K. Vandewal, and K. Leo, "Elementary steps in electrical doping of organic semiconductors," Nat. Commun. **9**, 1182 (2018).

[51] A. Privitera, G. Londi, M. Riede, G. D'Avino, and D. Beljonne, "Molecular quadrupole moments promote ground-state charge generation in doped organic semiconductors," Adv. Funct. Mater. **30**, 2004600 (2020).

[52] W. Kaiser, J. Popp, M. Rinderle, T. Albes, and A. Gagliardi, "Generalized kinetic monte carlo framework for organic electronics," Algorithms **11**, 37 (2018).

[53]H. Abdalla, G. Zuo, and M. Kemerink, "Range and energetics of charge hopping in organic semiconductors," Phys. Rev. B **96**, 241202 (2017).

[54]T. J. Aubry, J. C. Axtell, V. M. Basile, K. J. Winchell, J. R. Lindemuth, T. M. Porter, J.-Y. Liu, A. N. Alexandrova, C. P. Kubiak, S. H. Tolbert, A. M. Spokoyny, and B. J. Schwartz, "Dodecaborane-based dopants designed to shield anion electrostatics lead to increased carrier mobility in a doped conjugated polymer," Adv. Mater. **31**, 1805647 (2019).

[55]Y. Karpov, T. Erdmann, I. Raguzin, M. Al-Hussein, M. Binner, U. Lappan, M. Stamm, K. L. Gerasimov, T. Beryozkina, V. Bakulev, D. V. Anokhin, D. A. Ivanov, F. Günther, S. Gemming, G. Seifert, B. Voit, R. Di Pietro, and A. Kiriy, "High conductivity in molecularly p-doped diketopyrrolopyrrole-based polymer: The impact of a high dopant strength and good structural order," Adv. Mater. **28**, 6003-6010 (2016).

[56]Z. Liang, Y. Zhang, M. Souri, X. Luo, Alex M. Boehm, R. Li, Y. Zhang, T. Wang, D.-Y. Kim, J. Mei, S. R. Marder, and K. R. Graham, "Influence of dopant size and electron affinity on the electrical conductivity and thermoelectric properties of a series of conjugated polymers," J. Mater. Chem. A **6**, 16495-16505 (2018).

[57]J. Li, G. Zhang, D. M. Holm, I. E. Jacobs, B. Yin, P. Stroeve, M. Mascal, and A. J. Moulé, "Introducing solubility control for improved organic p-type opants," Chem. Mater. **27**, 5765-5774 (2015).

[58]T. J. Aubry, K. J. Winchell, C. Z. Salamat, V. M. Basile, J. R. Lindemuth, J. M. Stauber, J. C. Axtell, R. M. Kubena, M. D. Phan, M. J. Bird, A. M. Spokoyny, S. H. Tolbert, and B. J. Schwartz, "Tunable dopants with intrinsic counterion separation reveal the effects of electron affinity on dopant intercalation and free carrier production in sequentially doped conjugated polymer Films," Adv. Funct. Mater. **30**, 2001800 (2020).

[59]I. E. Jacobs, Y. Lin, Y. Huang, X. Ren, D. Simatos, C. Chen, D. Tjhe, M. Statz, L. Lai, P. A. Finn, W. G. Neal, G. D'Avino, V. Lemaur, S. Fratini, D. Beljonne, J. Strzalka, C. B. Nielsen, S. Barlow, S. R. Marder, I. McCulloch, and H. Sirringhaus, "High-efficiency ion-exchange doping of conducting polymers," Adv. Mater., 2102988 (2021).

[60]T. Z. Ma, B. X. Dong, J. W. Onorato, J. Niklas, O. Poluektov, C. K. Luscombe, and S. N. Patel, "Correlating conductivity and Seebeck coefficient to doping within crystalline and amorphous domains in poly(3-(methoxyethoxyethoxy)thiophene)," J. Polym. Sci. **59**, 2797-2808 (2021).

[61]M. L. Tietze, P. Pahner, K. Schmidt, K. Leo, and B. Lüssem, "Doped organic semiconductors: Trap-filling, impurity saturation, and reserve regimes," Adv. Funct. Mater. **25**, 2701-2707 (2015).

[62]V. I. Arkhipov, E. V. Emelianova, P. Heremans, and H. Bässler, "Analytic model of carrier mobility in doped disordered organic semiconductors," Phys. Rev. B **72**, 235202 (2005).

[63] V. I. Arkhipov, P. Heremans, E. V. Emelianova, and H. Bässler, "Effect of doping on the density-of-states distribution and carrier hopping in disordered organic semiconductors," Phys. Rev. B **71**, 045214 (2005).

[64] G. Zuo, H. Abdalla, and M. Kemerink, "Conjugated polymer blends for organic thermoelectrics," Adv. Electron. Mater. **5**, 1800821 (2019).

[65] Z. Liang, H. H. Choi, X. Luo, T. Liu, A. Abtahi, U. S. Ramasamy, J. A. Hitron, K. N. Baustert, J. L. Hempel, A. M. Boehm, A. Ansary, D. R. Strachan, J. Mei, C. Risko, V. Podzorov, and K. R. Graham, "n-type charge transport in heavily p-doped polymers," Nat. Mater. **20**, 518-524 (2021).

[66] K. Xu, T. P. Ruoko, M. Shokrani, D. Scheunemann, H. Abdalla, H. D. Sun, C. Y. Yang, Y. Puttisong, N. B. Kolhe, J. S. M. Figueroa, J. O. Pedersen, T. Ederth, W. M. Chen, M. Berggren, S. A. Jenekhe, D. Fazzi, M. Kemerink, and S. Fabiano, "On the Origin of Seebeck Coefficient Inversion in Highly Doped Conducting Polymers," Adv. Funct. Mater., 2112276 (2022).

[67] D. Derewjanko, D. Scheunemann, E. Järsvall, A. I. Hofmann, C. Müller, and M. Kemerink, "Delocalization Enhances Conductivity at High Doping Concentrations," Adv. Funct. Mater., 2112262 (2022).

[68] M. Lögdlund, R. Lazzaroni, S. Stafström, W. R. Salaneck, and J. L. Brédas, "Direct observation of charge-induced π-electronic structural changes in a conjugated polymer," Phys. Rev. Lett. **63**, 1841-1844 (1989).

[69] C. Gaul, S. Hutsch, M. Schwarze, K. S. Schellhammer, F. Bussolotti, S. Kera, G. Cuniberti, K. Leo, and F. Ortmann, "Insight into doping efficiency of organic semiconductors from the analysis of the density of states in n-doped $C_{60}$ and ZnPc," Nat. Mater. **17**, 439-444 (2018).

[70] X. Lin, G. E. Purdum, Y. Zhang, S. Barlow, S. R. Marder, Y.-L. Loo, and A. Kahn, "Impact of a low concentration of dopants on the distribution of gap states in a molecular semiconductor," Chem. Mater. **28**, 2677-2684 (2016).

[71] A. Fediai, F. Symalla, P. Friederich, and W. Wenzel, "Disorder compensation controls doping efficiency in organic semiconductors," Nat. Commun. **10**, 4547 (2019).

[72] P. Boufflet, G. Bovo, L. Occhi, H.-K. Yuan, Z. Fei, Y. Han, T. D. Anthopoulos, P. N. Stavrinou, and M. Heeney, "The influence of backbone fluorination on the dielectric constant of conjugated polythiophenes," Adv. Electron. Mater. **4**, 1700375 (2018).

[73] B. Xu, X. Yi, T.-Y. Huang, Z. Zheng, J. Zhang, A. Salehi, V. Coropceanu, C. H. Y. Ho, S. R. Marder, M. F. Toney, J.-L. Brédas, F. So, and J. R. Reynolds, "Donor conjugated polymers with polar side chain groups: The role of dielectric constant and energetic disorder on photovoltaic performance," Adv. Funct. Mater. **28**, 1803418 (2018).

[74] P. Yang, M. Yuan, D. F. Zeigler, S. E. Watkins, J. A. Lee, and C. K. Luscombe, "Influence of fluorine substituents on the film dielectric constant and open-circuit voltage in organic photovoltaics," J. Mater. Chem. C **2**, 3278-3284 (2014).

[75] Y. Lu, Z. Xiao, Y. Yuan, H. Wu, Z. An, Y. Hou, C. Gao, and J. Huang, "Fluorine substituted thiophene–quinoxaline copolymer to reduce the HOMO level and increase the dielectric constant for high open-circuit voltage organic solar cells," J. Mater. Chem. C **1**, 630-637 (2013).

[76] C. Wang, Z. Zhang, S. Pejić, R. Li, M. Fukuto, L. Zhu, and G. Sauvé, "High dielectric constant semiconducting poly(3-alkylthiophene)s from side chain modification with polar sulfinyl and sulfonyl groups," Macromolecules **51**, 9368-9381 (2018).

[77] N. Cho, C. W. Schlenker, K. M. Knesting, P. Koelsch, H.-L. Yip, D. S. Ginger, and A. K.-Y. Jen, "High-dielectric constant side-chain polymers show reduced non-geminate recombination in heterojunction solar cells," Adv. Energy Mater. **4**, 1301857 (2014).

[78] Y. Sun, S.-C. Chien, H.-L. Yip, K.-S. Chen, Y. Zhang, J. A. Davies, F.-C. Chen, B. Lin, and A. K. Y. Jen, "Improved thin film morphology and bulk-heterojunction solar cell performance through systematic tuning of the surface energy of conjugated polymers," J. Mater. Chem. **22**, 5587-5595 (2012).

[79] J. Brebels, J. V. Manca, L. Lutsen, D. Vanderzande, and W. Maes, "High dielectric constant conjugated materials for organic photovoltaics," J. Mater. Chem. A **5**, 24037-24050 (2017).

[80] X. Chen, Z. Zhang, Z. Ding, J. Liu, and L. Wang, "Diketopyrrolopyrrole-based conjugated polymers bearing branched oligo(ethylene glycol) side chains for photovoltaic devices," Angew. Chem. Int. Ed. **55**, 10376-10380 (2016).

[81] S. Torabi, F. Jahani, I. Van Severen, C. Kanimozhi, S. Patil, R. W. A. Havenith, R. C. Chiechi, L. Lutsen, D. J. M. Vanderzande, T. J. Cleij, J. C. Hummelen, and L. J. A. Koster, "Strategy for enhancing the dielectric constant of organic semiconductors without sacrificing charge carrier mobility and solubility," Adv. Funct. Mater. **25**, 150-157 (2015).

[82] M. Lenes, F. B. Kooistra, J. C. Hummelen, I. V. Severen, L. Lutsen, D. Vanderzande, T. J. Cleij, and P. W. M. Blom, "Charge dissociation in polymer:fullerene bulk heterojunction solar cells with enhanced permittivity," J. Appl. Phys. **104**, 114517 (2008).

[83] M. Breselge, I. Van Severen, L. Lutsen, P. Adriaensens, J. Manca, D. Vanderzande, and T. Cleij, "Comparison of the electrical characteristics of four 2,5-substituted poly(p-phenylene vinylene) derivatives with different side chains," Thin Solid Films **511-512**, 328-332 (2006).

[84] Note that p(g$_4$2T-T) has a lower side-chain grafting density and IE$_{osc}$ = 4.7 eV compared to regioregular P3HT with IE$_{osc}$ = 5.1 eV due to the electron-donating effect of the oxygen adjacent to some of the thiophene rings of p(g$_4$2T-T); IE$_{osc}$ = 5.1 eV + e × oxidation potential measured vs. ferrocene/ferrocenium, Fc/Fc$^+$.

[85]R. Kroon, D. Kiefer, D. Stegerer, L. Yu, M. Sommer, and C. Muller, "Polar side chains enhance processability, electrical conductivity, and thermal stability of a molecularly p-doped polythiophene," Adv. Mater. **29**, 1700930 (2017).

[86]I. E. Jacobs, E. W. Aasen, J. L. Oliveira, T. N. Fonseca, J. D. Roehling, J. Li, G. Zhang, M. P. Augustine, M. Mascal, and A. J. Moulé, "Comparison of solution-mixed and sequentially processed P3HT:F$_4$TCNQ films: effect of doping-induced aggregation on film morphology," J. Mater. Chem. C **4**, 3454-3466 (2016).

[87]J. Hynynen, D. Kiefer, and C. Müller, "Influence of crystallinity on the thermoelectric power factor of P3HT vapour-doped with F$_4$TCNQ," RSC Adv. **8**, 1593-1599 (2018).

[88]J. Hynynen, D. Kiefer, L. Yu, R. Kroon, R. Munir, A. Amassian, M. Kemerink, and C. Müller, "Enhanced electrical conductivity of molecularly p-doped poly(3-hexylthiophene) through understanding the correlation with solid-state order," Macromolecules **50**, 8140-8148 (2017).

[89]A. Hamidi-Sakr, L. Biniek, J.-L. Bantignies, D. Maurin, L. Herrmann, N. Leclerc, P. Lévêque, V. Vijayakumar, N. Zimmermann, and M. Brinkmann, "A versatile method to fabricate highly in-plane aligned conducting polymer films with anisotropic charge transport and thermoelectric properties: The key role of alkyl side chain layers on the doping mechanism," Adv. Funct. Mater. **27**, 1700173 (2017).

[90]E. Lim, K. A. Peterson, G. M. Su, and M. L. Chabinyc, "Thermoelectric properties of poly(3-hexylthiophene) (P3HT) doped with 2,3,5,6-tetrafluoro-7,7,8,8-tetracyanoquinodimethane (F$_4$TCNQ) by vapor-phase infiltration," Chem. Mater. **30**, 998-1010 (2018).

[91]K. Xu, H. Sun, T.-P. Ruoko, G. Wang, R. Kroon, N. B. Kolhe, Y. Puttisong, X. Liu, D. Fazzi, K. Shibata, C.-Y. Yang, N. Sun, G. Persson, A. B. Yankovich, E. Olsson, H. Yoshida, W. M. Chen, M. Fahlman, M. Kemerink, S. A. Jenekhe, C. Müller, M. Berggren, and S. Fabiano, "Ground-state electron transfer in all-polymer donor–acceptor heterojunctions," Nat. Mater. **19**, 738-744 (2020).

[92]C. Wang, D. T. Duong, K. Vandewal, J. Rivnay, and A. Salleo, "Optical measurement of doping efficiency in poly(3-hexylthiophene) solutions and thin films," Phys. Rev. B **91**, 085205 (2015).

[93]T. F. Harrelson, Y. Q. Cheng, J. Li, I. E. Jacobs, A. J. Ramirez-Cuesta, R. Faller, and A. J. Moulé, "Identifying atomic scale structure in undoped/doped semicrystalline P3HT using inelastic neutron scattering," Macromolecules **50**, 2424-2435 (2017).

[94]D. T. Scholes, S. A. Hawks, P. Y. Yee, H. Wu, J. R. Lindemuth, S. H. Tolbert, and B. J. Schwartz, "Overcoming film quality issues for conjugated polymers doped with F$_4$TCNQ by solution sequential processing: Hall Effect, structural, and optical measurements," J. Phys. Chem. Lett. **6**, 4786-4793 (2015).

[95] L. Müller, D. Nanova, T. Glaser, S. Beck, A. Pucci, A. K. Kast, R. R. Schröder, E. Mankel, P. Pingel, D. Neher, W. Kowalsky, and R. Lovrincic, "Charge-transfer–solvent interaction predefines doping efficiency in p-doped P3HT films," Chem. Mater. **28**, 4432-4439 (2016).

[96] D. Scheunemann, and M. Kemerink, in *Organic flexible electronics*, (Eds: P. Cosseddu, M. Caironi), Woodhead Publishing, 2021, 165-197.

[97] M. Comin, S. Fratini, X. Blase, and G. D'Avino, "Doping-Induced Dielectric Catastrophe Prompts Free-Carrier Release in Organic Semiconductors," Adv. Mater. **34**, 2105376 (2022).

[98] M. Koopmans, M. A. T. Leiviskä, J. Liu, J. Dong, L. Qiu, J. C. Hummelen, G. Portale, M. C. Heiber, and L. J. A. Koster, "Electrical conductivity of doped organic semiconductors limited by carrier–carrier interactions," ACS Appl. Mater. Interfaces **12**, 56222-56230 (2020).

[99] E. M. Thomas, K. A. Peterson, A. H. Balzer, D. Rawlings, N. Stingelin, R. A. Segalman, and M. L. Chabinyc, "Effects of Counter-Ion Size on Delocalization of Carriers and Stability of Doped Semiconducting Polymers," Adv. Electron. Mater. **6**, 2000595 (2020).

[100] I. E. Jacobs, G. D'Avino, V. Lemaur, Y. Lin, Y. Huang, C. Chen, T. F. Harrelson, W. Wood, L. J. Spalek, T. Mustafa, C. A. O'Keefe, X. Ren, D. Simatos, D. Tjhe, M. Statz, J. W. Strzalka, J.-K. Lee, I. McCulloch, S. Fratini, D. Beljonne, and H. Sirringhaus, "Structural and Dynamic Disorder, Not Ionic Trapping, Controls Charge Transport in Highly Doped Conducting Polymers," J. Am. Chem. Soc. **144**, 3005-3019 (2022).

[101] D. Kim, and I. Zozoulenko, "Why is pristine PEDOT oxidized to 33%? A density functional theory study of oxidative polymerization mechanism," J. Phys. Chem. B **123**, 5160-5167 (2019).

[102] V. Untilova, J. Hynynen, A. I. Hofmann, D. Scheunemann, Y. Zhang, S. Barlow, M. Kemerink, S. R. Marder, L. Biniek, C. Müller, and M. Brinkmann, "High thermoelectric power factor of poly(3-hexylthiophene) through in-plane alignment and doping with a molybdenum dithiolene complex," Macromolecules **53**, 6314-6321 (2020).

[103] D. T. Duong, C. Wang, E. Antono, M. F. Toney, and A. Salleo, "The chemical and structural origin of efficient p-type doping in P3HT," Org. Electron. **14**, 1330-1336 (2013).

[104] J. E. Cochran, M. J. N. Junk, A. M. Glaudell, P. L. Miller, J. S. Cowart, M. F. Toney, C. J. Hawker, B. F. Chmelka, and M. L. Chabinyc, "Molecular interactions and ordering in electrically doped polymers: Blends of PBTTT and $F_4$TCNQ," Macromolecules **47**, 6836-6846 (2014).

[105] P. Y. Yee, D. T. Scholes, B. J. Schwartz, and S. H. Tolbert, "Dopant-induced ordering of amorphous regions in regiorandom P3HT," J. Phys. Chem. Lett. **10**, 4929-4934 (2019).

[106] W. Liu, L. Müller, S. Ma, S. Barlow, S. R. Marder, W. Kowalsky, A. Köhn, and R. Lovrincic, "Origin of the π–π spacing change upon doping of semiconducting polymers," J. Phys. Chem. C **122**, 27983-27990 (2018).

[107]J. Li, C. Koshnick, S. O. Diallo, S. Ackling, D. M. Huang, I. E. Jacobs, T. F. Harrelson, K. Hong, G. Zhang, J. Beckett, M. Mascal, and A. J. Moulé, "Quantitative measurements of the temperature-dependent microscopic and macroscopic dynamics of a molecular dopant in a conjugated polymer," Macromolecules **50**, 5476-5489 (2017).

[108]I. E. Jacobs, E. W. Aasen, D. Nowak, J. Li, W. Morrison, J. D. Roehling, M. P. Augustine, and A. J. Moulé, "Direct-write optical patterning of P3HT films beyond the diffraction limit," Adv. Mater. **29**, 1603221 (2017).

[109]V. Vijayakumar, P. Durand, H. Zeng, V. Untilova, L. Herrmann, P. Algayer, N. Leclerc, and M. Brinkmann, "Influence of dopant size and doping method on the structure and thermoelectric properties of PBTTT films doped with $F_6$TCNNQ and $F_4$TCNQ," J. Mater. Chem. C **8**, 16470-16482 (2020).

[110]C. J. Boyle, M. Upadhyaya, P. Wang, L. A. Renna, M. Lu-Díaz, S. Pyo Jeong, N. Hight-Huf, L. Korugic-Karasz, M. D. Barnes, Z. Aksamija, and D. Venkataraman, "Tuning charge transport dynamics via clustering of doping in organic semiconductor thin films," Nat. Commun. **10**, 2827 (2019).

[111]J. Liu, L. Qiu, R. Alessandri, X. Qiu, G. Portale, J. Dong, W. Talsma, G. Ye, A. A. Sengrian, P. C. T. Souza, M. A. Loi, R. C. Chiechi, S. J. Marrink, J. C. Hummelen, and L. J. A. Koster, "Enhancing molecular n-type doping of donor–acceptor copolymers by tailoring side chains," Adv. Mater. **30**, 1704630 (2018).

[112]J. Liu, L. Qiu, G. Portale, M. Koopmans, G. ten Brink, J. C. Hummelen, and L. J. A. Koster, "N-Type organic thermoelectrics: Improved power factor by tailoring host–dopant miscibility," Adv. Mater. **29**, 1701641 (2017).

[113]D. Kiefer, A. Giovannitti, H. Sun, T. Biskup, A. Hofmann, M. Koopmans, C. Cendra, S. Weber, L. J. Anton Koster, E. Olsson, J. Rivnay, S. Fabiano, I. McCulloch, and C. Müller, "Enhanced n-doping efficiency of a naphthalenediimide-based copolymer through polar side chains for organic thermoelectrics," ACS Energy Lett. **3**, 278-285 (2018).

[114]B. X. Dong, C. Nowak, J. W. Onorato, T. Ma, J. Niklas, O. G. Poluektov, G. Grocke, M. F. DiTusa, F. A. Escobedo, C. K. Luscombe, P. F. Nealey, and S. N. Patel, "Complex relationship between side-chain polarity, conductivity, and thermal stability in molecularly doped conjugated polymers," Chem. Mater. **33**, 741-753 (2021).

[115]S. Kohno, Y. Yamashita, N. Kasuya, T. Mikie, I. Osaka, K. Takimiya, J. Takeya, and S. Watanabe, "Controlled steric selectivity in molecular doping towards closest-packed supramolecular conductors," Commun. Mater. **1**, 79 (2020).

[116] H. Li, M. E. DeCoster, C. Ming, M. Wang, Y. Chen, P. E. Hopkins, L. Chen, and H. E. Katz, "Enhanced molecular doping for high conductivity in polymers with volume freed for dopants," Macromolecules **52**, 9804-9812 (2019).

[117] M. R. Andersson, O. Thomas, W. Mammo, M. Svensson, M. Theander, and O. Inganäs, "Substituted polythiophenes designed for optoelectronic devices and conductors," J. Mater. Chem. **9**, 1933-1940 (1999).

[118] V. Vijayakumar, E. Zaborova, L. Biniek, H. Zeng, L. Herrmann, A. Carvalho, O. Boyron, N. Leclerc, and M. Brinkmann, "Effect of alkyl side chain length on doping kinetics, thermopower, and charge transport properties in highly oriented $F_4TCNQ$-doped PBTTT films," ACS Appl. Mater. Interfaces **11**, 4942-4953 (2019).

[119] J. Liu, G. Ye, H. G. O. Potgieser, M. Koopmans, S. Sami, M. I. Nugraha, D. R. Villalva, H. Sun, J. Dong, X. Yang, X. Qiu, C. Yao, G. Portale, S. Fabiano, T. D. Anthopoulos, D. Baran, R. W. A. Havenith, R. C. Chiechi, and L. J. A. Koster, "Amphipathic side chain of a conjugated polymer optimizes dopant location toward efficient n-type organic thermoelectrics," Adv. Mater. **33**, 2006694 (2021).

[120] D. Di Nuzzo, C. Fontanesi, R. Jones, S. Allard, I. Dumsch, U. Scherf, E. von Hauff, S. Schumacher, and E. Da Como, "How intermolecular geometrical disorder affects the molecular doping of donor–acceptor copolymers," Nat. Commun. **6**, 6460 (2015).

[121] E. Lim, A. M. Glaudell, R. Miller, and M. L. Chabinyc, "The role of ordering on the thermoelectric properties of blends of regioregular and regiorandom poly(3-hexylthiophene)," Adv. Electron. Mater. **5**, 1800915 (2019).

[122] A. R. Chew, R. Ghosh, Z. Shang, F. C. Spano, and A. Salleo, "Sequential doping reveals the importance of amorphous chain rigidity in charge transport of semi-crystalline polymers," J. Phys. Chem. Lett. **8**, 4974-4980 (2017).

[123] M. T. Fontana, D. A. Stanfield, D. T. Scholes, K. J. Winchell, S. H. Tolbert, and B. J. Schwartz, "Evaporation vs solution sequential doping of conjugated polymers: $F_4TCNQ$ doping of micrometer-thick P3HT films for thermoelectrics," J. Phys. Chem. C **123**, 22711-22724 (2019).

[124] L. Yu, D. Scheunemann, A. Lund, D. Kiefer, and C. Müller, "Sequential doping of solid chunks of a conjugated polymer for body-heat-powered thermoelectric modules," Appl. Phys. Lett. **119**, 181902 (2021).

[125] J. Hynynen, E. Järsvall, R. Kroon, Y. Zhang, S. Barlow, S. R. Marder, M. Kemerink, A. Lund, and C. Müller, "Enhanced thermoelectric power factor of tensile drawn poly(3-hexylthiophene)," ACS Macro Lett. **8**, 70-76 (2019).

[126]S. N. Patel, A. M. Glaudell, K. A. Peterson, E. M. Thomas, K. A. O'Hara, E. Lim, and M. L. Chabinyc, "Morphology controls the thermoelectric power factor of a doped semiconducting polymer," Sci. Adv. **3**, e1700434 (2017).

[127]R. Kroon, J. D. Ryan, D. Kiefer, L. Yu, J. Hynynen, E. Olsson, and C. Müller, "Bulk doping of millimeter-thick conjugated polymer foams for plastic thermoelectrics," Adv. Funct. Mater. **27**, 1704183 (2017).

[128]C. Müller, C. P. Radano, P. Smith, and N. Stingelin-Stutzmann, "Crystalline–crystalline poly(3-hexylthiophene)–polyethylene diblock copolymers: Solidification from the melt," Polymer **49**, 3973-3978 (2008).

[129]D. Beretta, A. Perego, G. Lanzani, and M. Caironi, "Organic flexible thermoelectric generators: from modeling, a roadmap towards applications," Sustain. Energy & Fuels **1**, 174-190 (2017).

[130]A. Lund, Y. Tian, S. Darabi, and C. Müller, "A polymer-based textile thermoelectric generator for wearable energy harvesting," J.Power Sources **480**, 228836 (2020).

[131]J. Li, C. W. Rochester, I. E. Jacobs, S. Friedrich, P. Stroeve, M. Riede, and A. J. Moulé, "Measurement of small molecular dopant $F_4TCNQ$ and $C_{60}F_{36}$ diffusion in organic bilayer architectures," ACS Appl. Mater. Interfaces **7**, 28420-28428 (2015).

[132]J. Li, C. W. Rochester, I. E. Jacobs, E. W. Aasen, S. Friedrich, P. Stroeve, and A. J. Moule, "The effect of thermal annealing on dopant site choice in conjugated polymers," Org. Electron. **33**, 23-31 (2016).

[133]K. E. Watts, B. Neelamraju, M. Moser, I. McCulloch, E. L. Ratcliff, and J. E. Pemberton, "Thermally induced formation of $HF_4TCNQ^-$ in $F_4TCNQ$-doped regioregular P3HT," J. Phys. Chem. Lett. **11**, 6586-6592 (2020).

[134]K. E. Watts, B. Neelamraju, E. L. Ratcliff, and J. E. Pemberton, "Stability of charge transfer states in $F_4TCNQ$-doped P3HT," Chem. Mater. **31**, 6986-6994 (2019).

[135]D. Nava, Y. Shin, M. Massetti, X. Jiao, T. Biskup, M. S. Jagadeesh, A. Calloni, L. Duò, G. Lanzani, C. R. McNeill, M. Sommer, and M. Caironi, "Drastic improvement of air stability in an n-type doped naphthalene-diimide polymer by thionation," ACS Appl. Energy Mater. **1**, 4626-4634 (2018).

[136]S. Wang, T.-P. Ruoko, G. Wang, S. Riera-Galindo, S. Hultmark, Y. Puttisong, F. Moro, H. Yan, W. M. Chen, M. Berggren, C. Müller, and S. Fabiano, "Sequential doping of ladder-type conjugated polymers for thermally stable n-type organic conductors," ACS Appl. Mater. Interfaces **12**, 53003-53011 (2020).

[137]C.-Y. Yang, Y.-F. Ding, D. Huang, J. Wang, Z.-F. Yao, C.-X. Huang, Y. Lu, H.-I. Un, F.-D. Zhuang, J.-H. Dou, C.-a. Di, D. Zhu, J.-Y. Wang, T. Lei, and J. Pei, "A thermally activated and highly miscible dopant for n-type organic thermoelectrics," Nat. Commun. **11**, 3292 (2020).

[138]C.-Y. Yang, M.-A. Stoeckel, T.-P. Ruoko, H.-Y. Wu, X. Liu, N. B. Kolhe, Z. Wu, Y. Puttisong, C. Musumeci, M. Massetti, H. Sun, K. Xu, D. Tu, W. M. Chen, H. Y. Woo, M. Fahlman, S. A. Jenekhe, M.

Berggren, and S. Fabiano, "A high-conductivity n-type polymeric ink for printed electronics," Nat. Commun. **12**, 2354 (2021).

[139]Y. Lu, Z. D. Yu, H. I. Un, Z. F. Yao, H. Y. You, W. Jin, L. Li, Z. Y. Wang, B. W. Dong, S. Barlow, E. Longhi, C. A. Di, D. Zhu, J. Y. Wang, C. Silva, S. R. Marder, and J. Pei, "Persistent conjugated backbone and disordered lamellar packing impart polymers with efficient n-doping and high conductivities," Adv. Mater. **33**, 2005946 (2021).

[140]J. Liu, G. Ye, B. V. Zee, J. Dong, X. Qiu, Y. Liu, G. Portale, R. C. Chiechi, and L. J. A. Koster, "N-Type organic thermoelectrics of donor-acceptor copolymers: Improved power factor by molecular tailoring of the density of states," Adv. Mater. **30**, 1804290 (2018).

[141]E. M. Thomas, B. C. Popere, H. Fang, M. L. Chabinyc, and R. A. Segalman, "Role of disorder induced by doping on the thermoelectric properties of semiconducting polymers," Chem. Mater. **30**, 2965-2972 (2018).

[142]G. Zuo, X. Liu, M. Fahlman, and M. Kemerink, "High seebeck coefficient in mixtures of conjugated polymers," Adv. Funct. Mater. **28**, 1703280 (2018).

[143]A. Abtahi, S. Johnson, S. M. Park, X. Luo, Z. Liang, J. Mei, and K. R. Graham, "Designing π-conjugated polymer blends with improved thermoelectric power factors," J. Mater. Chem. A **7**, 19774-19785 (2019).

[144]R. Sarabia-Riquelme, M. Shahi, J. W. Brill, and M. C. Weisenberger, "Effect of drawing on the electrical, thermoelectrical, and mechanical properties of wet-spun PEDOT:PSS fibers," ACS Appl. Polym. Mater. **1**, 2157-2167 (2019).

[145]Q. Wei, M. Mukaida, K. Kirihara, and T. Ishida, "Experimental studies on the anisotropic thermoelectric properties of conducting polymer films," ACS Macro Lett. **3**, 948-952 (2014).

[146]Y. Kim, A. Lund, H. Noh, A. I. Hofmann, M. Craighero, S. Darabi, S. Zokaei, J. I. Park, M.-H. Yoon, and C. Müller, "Robust PEDOT:PSS wet-spun fibers for thermoelectric textiles," Macromol. Mater. Eng. **305**, 1900749 (2020).

[147]D. Scheunemann, V. Vijayakumar, H. Zeng, P. Durand, N. Leclerc, M. Brinkmann, and M. Kemerink, "Rubbing and drawing: Generic ways to improve the thermoelectric power factor of organic semiconductors?," Adv. Electron. Mater. **6**, (2020).

[148]V. Untilova, T. Biskup, L. Biniek, V. Vijayakumar, and M. Brinkmann, "Control of chain alignment and crystallization helps enhance charge conductivities and thermoelectric power factors in sequentially doped P3HT:F$_4$TCNQ films," Macromolecules **53**, 2441-2453 (2020).

[149]S. Hwang, W. J. Potscavage, Y. S. Yang, I. S. Park, T. Matsushima, and C. Adachi, "Solution-processed organic thermoelectric materials exhibiting doping-concentration-dependent polarity," Phys. Chem. Chem. Phys. **18**, 29199-29207 (2016).

[150]C. Y. Yang, W. L. Jin, J. Wang, Y. F. Ding, S. Nong, K. Shi, Y. Lu, Y. Z. Dai, F. D. Zhuang, T. Lei, C. A. Di, D. Zhu, J. Y. Wang, and J. Pei, "Enhancing the n-type conductivity and thermoelectric performance of donor-acceptor copolymers through donor engineering," Adv. Mater. **30**, e1802850 (2018).

[151]H. Zeng, M. Mohammed, V. Untilova, O. Boyron, N. Berton, P. Limelette, B. Schmaltz, and M. Brinkmann, "Fabrication of oriented n-type thermoelectric polymers by polarity switching in a DPP-based donor–acceptor copolymer doped with $FeCl_3$," Adv. Electron. Mater. **7**, (2021).

[152]J. M. Ziman, *Principles of the theory of solids*, Cambridge University Press, Cambridge 1972.

[153]R. Franz, and G. Wiedemann, "Ueber die Wärme-Leitungsfähigkeit der Metalle," Annalen der Physik **165**, 497-531 (1853).

[154]J. Liu, X. Wang, D. Li, N. E. Coates, R. A. Segalman, and D. G. Cahill, "Thermal conductivity and elastic constants of PEDOT:PSS with high electrical conductivity," Macromolecules **48**, 585-591 (2015).

[155]A. Weathers, Z. U. Khan, R. Brooke, D. Evans, M. T. Pettes, J. W. Andreasen, X. Crispin, and L. Shi, "Significant electronic thermal transport in the conducting polymer poly(3,4-ethylenedioxythiophene)," Adv. Mater. **27**, 2101-2106 (2015).

[156]M. B. Salamon, J. W. Bray, G. DePasquali, R. A. Craven, G. Stucky, and A. Schultz, "Thermal conductivity of tetrathiafulvalene-tetracyanoquinodimethane (TTF-TCNQ) near the metal-insulator transition," Phys. Rev. B **11**, 619-622 (1975).

[157]A. Miller, and E. Abrahams, "Impurity conduction at low concentrations," Phys. Rev. **120**, 745-755 (1960).

[158]G. T. Craven, and A. Nitzan, "Wiedemann–Franz law for molecular hopping transport," Nano Lett. **20**, 989-993 (2020).

[159]D. Scheunemann, and M. Kemerink, "Non-Wiedemann-Franz behavior of the thermal conductivity of organic semiconductors," Phys. Rev. B **101**, 075206 (2020).

[160]R. Schmechel, "Hopping transport in doped organic semiconductors: A theoretical approach and its application to p-doped zinc-phthalocyanine," J. Appl. Phys. **93**, 4653-4660 (2003).

[161]S. Ihnatsenka, X. Crispin, and I. V. Zozoulenko, "Understanding hopping transport and thermoelectric properties of conducting polymers," Phys. Rev. B **92**, 035201 (2015).

[162]O. Zapata-Arteaga, A. Perevedentsev, S. Marina, J. Martin, J. S. Reparaz, and M. Campoy-Quiles, "Reduction of the lattice thermal conductivity of polymer semiconductors by molecular doping," ACS Energy Lett. **5**, 2972-2978 (2020).

[163]X. Wang, C. D. Liman, N. D. Treat, M. L. Chabinyc, and D. G. Cahill, "Ultralow thermal conductivity of fullerene derivatives," Phys. Rev. B **88**, 075310 (2013).

[164]J. Liu, B. van der Zee, R. Alessandri, S. Sami, J. Dong, M. I. Nugraha, A. J. Barker, S. Rousseva, L. Qiu, X. Qiu, N. Klasen, R. C. Chiechi, D. Baran, M. Caironi, T. D. Anthopoulos, G. Portale, R. W. A. Havenith,

S. J. Marrink, J. C. Hummelen, and L. J. A. Koster, "N-type organic thermoelectrics: demonstration of ZT > 0.3," Nat. Commun. **11**, 5694 (2020).

[165]D. Beretta, A. J. Barker, I. Maqueira-Albo, A. Calloni, G. Bussetti, G. Dell'Erba, A. Luzio, L. Duò, A. Petrozza, G. Lanzani, and M. Caironi, "Thermoelectric properties of highly conductive Poly(3,4-ethylenedioxythiophene) polystyrene sulfonate printed thin films," ACS Appl. Mater. Interfaces **9**, 18151-18160 (2017).

[166]T. Degousée, V. Untilova, V. Vijayakumar, X. Xu, Y. Sun, M. Palma, M. Brinkmann, L. Biniek, and O. Fenwick, "High thermal conductivity states and enhanced figure of merit in aligned polymer thermoelectric materials," J. Mater. Chem. A **9**, 16065-16075 (2021).